\documentclass[acmsmall, dvipsnames,screen,nonacm]{acmart}

\usepackage{times,color}
\usepackage{listings}

\usepackage{amsthm,amssymb}
\usepackage{dsfont}
\usepackage{paralist}
\usepackage{graphicx}
\usepackage[ruled,vlined,linesnumbered]{algorithm2e}
\usepackage{wrapfig}
\usepackage{listings}
\usepackage{subcaption}
\usepackage{placeins}
\usepackage{float}
\usepackage{booktabs}
\usepackage{hyperref}
\usepackage{pdfpages}
\usepackage[parfill]{parskip}
\usepackage{subcaption}
\usepackage{boldline}
\usepackage{makecell}
\usepackage{cleveref}
\usepackage{wasysym}
\usepackage[thicklines]{cancel}
\usepackage{tikz}
\usepackage{pgfplots, pgfplotstable}

\usepackage{ifthen}

\newcommand{\DONE}[1]{}
\newcommand{\COMMENT}[1]{}

\newcommand{\ignore}[1]{}

\newcounter{programlinenumber}

\newboolean{TR}
\setboolean{TR}{false}
\ifthenelse{\boolean{TR}}{

\newcommand{\TrOnly}[1]{#1}
\newcommand{\SubOnly}[1]{}
\newcommand{\TrOnlyInFootnote}[1]{#1}
\newcommand{\TrOnlyInTable}[1]{#1}}
{

\newcommand{\TrOnly}[1]{}
\newcommand{\SubOnly}[1]{#1}
\newcommand{\TrOnlyInFootnote}[1]{}
\newcommand{\TrOnlyInTable}[1]{}}

\newcommand{\redone}{\textbf{\textcolor{Maroon}{\raisebox{.5pt}{\textcircled{\raisebox{-.9pt} {1}}}}}}

\newcommand{\purpletwo}{\textbf{\textcolor{Purple}{\raisebox{.5pt}{ \textcircled{\raisebox{-.9pt} {2}}}}}}

\newcommand{\greenthree}{\textbf{\textcolor{Green}{\raisebox{.5pt}{ \textcircled{\raisebox{-.9pt} {3}}}}}}

\newcommand{\hiddentext}[1]{}

\newcommand{\scode}[1]{{\small \texttt{#1}}}
\newcommand{\sname}[1]{{\small \textsc{#1}}}

\renewcommand{\phi}{\varphi}

\usepackage{color}
\definecolor{lightgray}{rgb}{.9,.9,.9}
\definecolor{darkgray}{rgb}{.4,.4,.4}
\definecolor{purple}{rgb}{0.65, 0.12, 0.82}
\lstdefinelanguage{JavaScript}{
  keywords={break, case, catch, continue, debugger, default, delete, do, else, false, finally, for, function, if, in, instanceof, new, null, return, switch, this, throw, true, try, typeof, var, void, while, with},
  morecomment=[l]{//},
  morecomment=[s]{/*}{*/},
  morestring=[b]',
  morestring=[b]",
  ndkeywords={class, export, boolean, throw, implements, import, this},
  keywordstyle=\color{blue}\bfseries,
  ndkeywordstyle=\color{darkgray}\bfseries,
  identifierstyle=\color{black},
  commentstyle=\color{purple}\ttfamily,
  stringstyle=\color{red}\ttfamily,
  sensitive=true
}

\newcommand{\uri}[1]{{\color{blue}Uri:[#1]}}
\newcommand{\ey}[1]{{\color{purple}Eran:[#1]}}
\newcommand{\shaked}[1]{{\color{red}Shaked:[#1]}}

\renewcommand{\shaked}[1]{}
\renewcommand{\uri}[1]{}
\renewcommand{\ey}[1]{}

\newcommand{\ctc}{{\small C$^3$}}
\newcommand{\cbefore}{\mathcal{C}}
\newcommand{\cafter}{\mathcal{C'}}
\newcommand{\pbefore}{\mathcal{P}}
\newcommand{\pafter}{\mathcal{P'}}
\newcommand{\ourtask}{\sname{EditCompletion}}

\setcopyright{none}
\acmPrice{}
\acmDOI{}
\acmYear{}
\copyrightyear{}
\acmSubmissionID{}
\acmJournal{PACMPL}
\acmVolume{4}
\acmNumber{OOPSLA}
\acmArticle{}
\acmMonth{}

\begin{document}

\title[]{A Structural Model for Contextual Code Changes}

\author{Shaked Brody}
\affiliation{
  \institution{Technion}            %
  \country{Israel}                    %
}
\email{shakedbr@cs.technion.ac.il}          %

\author{Uri Alon}
\affiliation{
  \institution{Technion}            %
  \country{Israel}                    %
}
\email{urialon@cs.technion.ac.il}          %

\author{Eran Yahav}
\affiliation{
  \institution{Technion}            %
  \country{Israel}                    %
}
\email{yahave@cs.technion.ac.il}          %

\begin{abstract}
We address the problem of predicting \emph{edit completions} based on a learned model that was trained on past edits. Given a code snippet that is partially edited, our goal is to predict a \emph{completion of the edit for the rest of the snippet}. We refer to this task as the \ourtask{} task and present a novel approach for tackling it. 
The main idea is to directly represent structural edits. This allows us to model the likelihood of the edit itself, rather than learning the likelihood of the edited code. We represent an edit operation as a path in the program's Abstract Syntax Tree (AST), originating from the source of the edit to the target of the edit. Using this representation, we present a powerful and lightweight neural model for the \ourtask{} task.

We conduct a thorough evaluation, comparing our approach to a variety of representation and modeling approaches that are driven by multiple strong models such as LSTMs, Transformers, and neural CRFs.
Our experiments show that our model achieves a 28\% relative gain over state-of-the-art sequential models and $2\times$ higher accuracy than syntactic models that learn to generate the edited \emph{code}, as opposed to modeling the \emph{edits} directly. 

Our code, dataset, and trained models are publicly available at \url{https://github.com/tech-srl/c3po/} .

\end{abstract}



\maketitle

\section{Introduction}\label{Se:Intro}
\begin{figure}
\begin{subfigure}[b]{1\textwidth}
  \centering
  \includegraphics[scale=0.44]{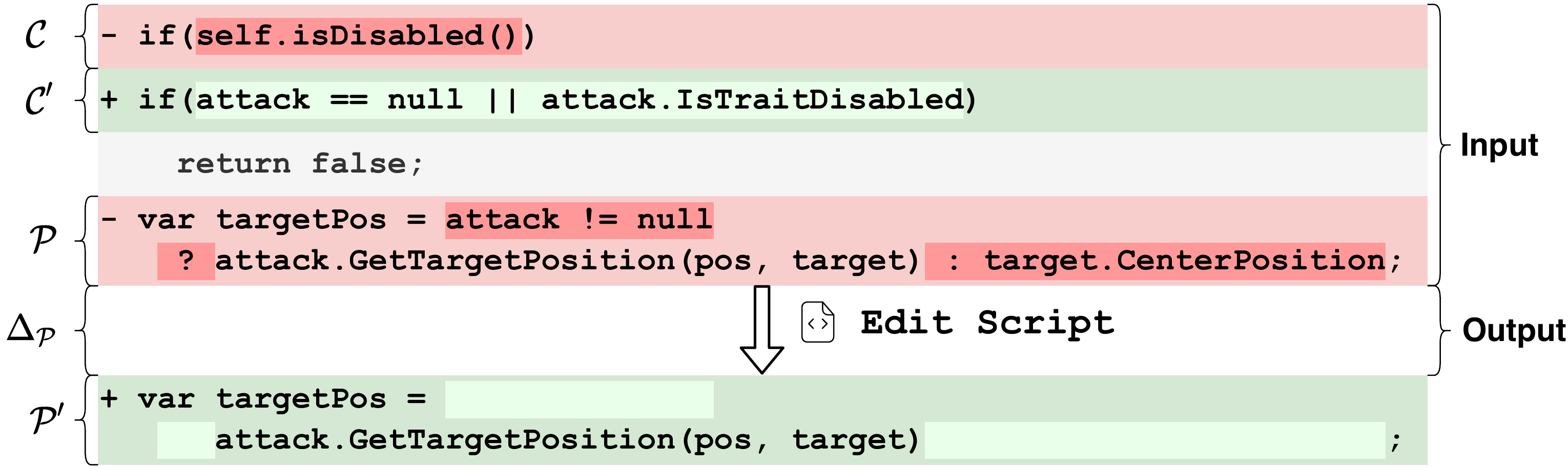}  
  \caption{The predicate of the \scode{if} statement in $\cbefore$ was edited to include a \scode{null} check for \scode{attack}. Thus, in $\pbefore$, the checking of \scode{attack != null} and the ternary operator can be removed. }
  \label{Fi:example_a}
\end{subfigure}

\par\bigskip 

\begin{subfigure}[b]{1\textwidth}
  \centering
  \includegraphics[scale=0.44]{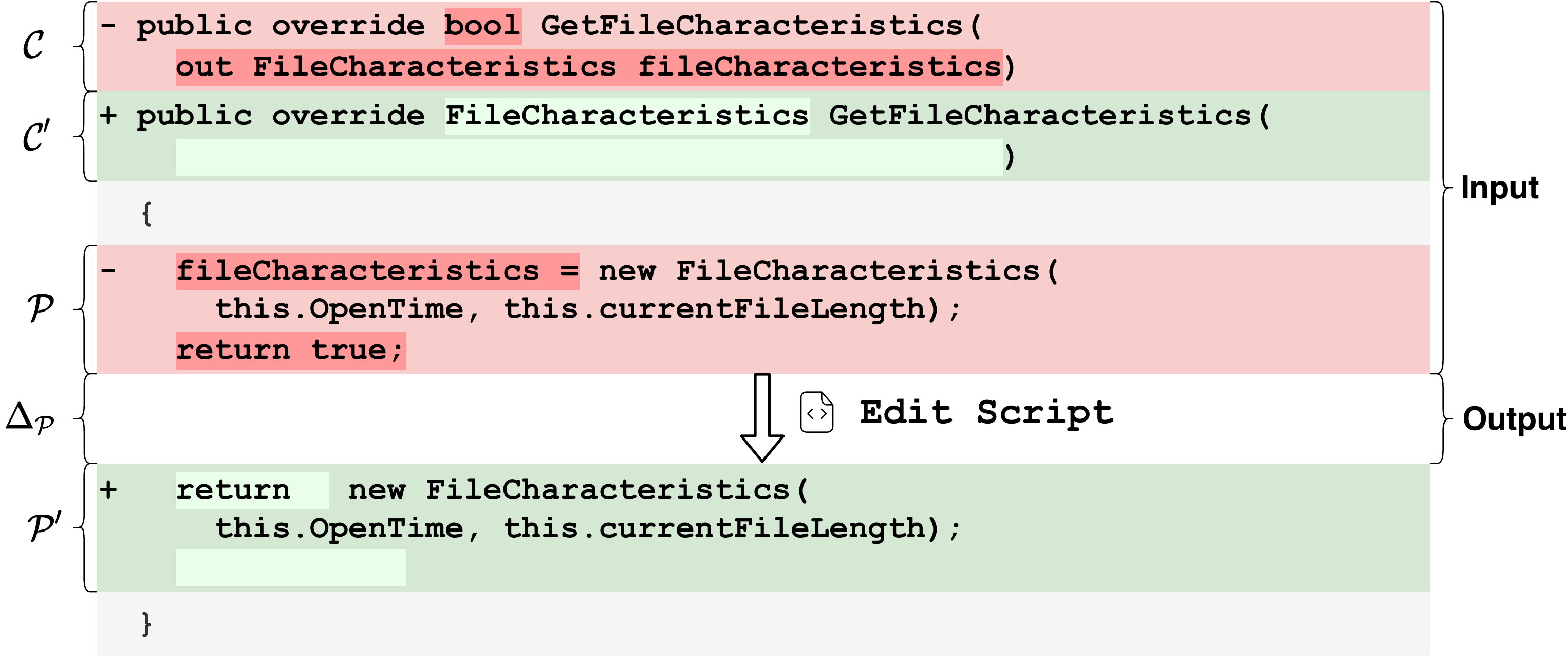}  
  \caption{The signature of \scode{GetFileCharacteristics} in $\cbefore$ was edited to return a \scode{FileCharacteristic} object instead of modifying an output parameter. 
  Thus, in $\pbefore$, the method should return a \scode{FileCharacteristic} object instead of returning \scode{true}. }
  \label{Fi:example_b}
\end{subfigure}
\caption{
Examples of \ourtask{}.
The input consists of a program fragment $\pbefore$ and edits that occurred in the context that transformed $\cbefore$ into $\cafter$. The output is $\Delta_\mathcal{P}$ -- an edit script that describes the likely edit. Applying  $\Delta_\mathcal{P}$ to $\pbefore$ results in $\pafter$ -- the code after the edit.
}
\label{Fi:example}
\end{figure}

\begin{figure}[t]
\begin{subfigure}[b]{.49\textwidth}
  \centering
  \includegraphics[scale=0.415]{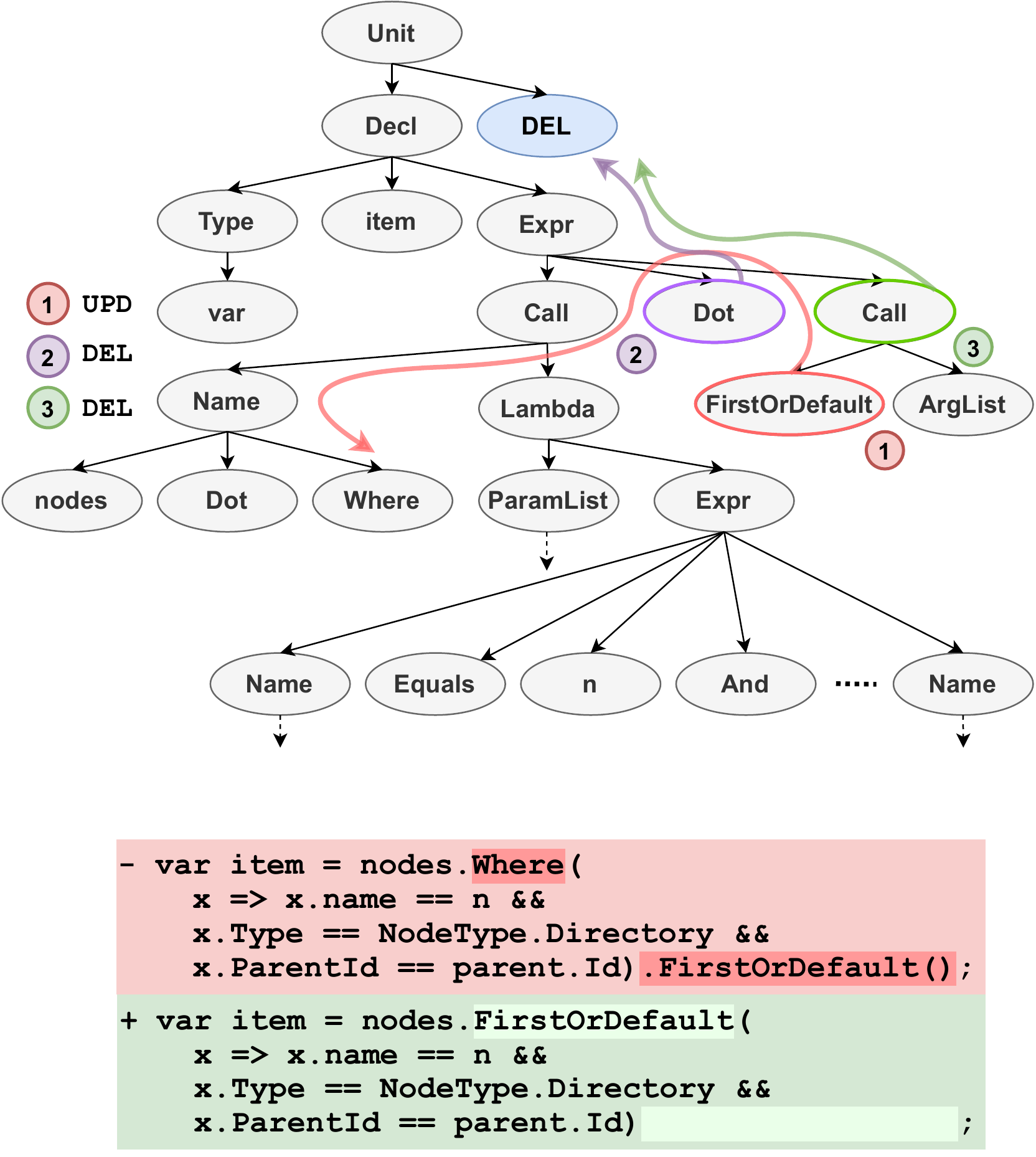}  
  \caption{}
  \label{Fi:gen_fig_a}
\end{subfigure}
\begin{subfigure}[b]{.49\textwidth}
  \centering
  \includegraphics[scale=0.415]{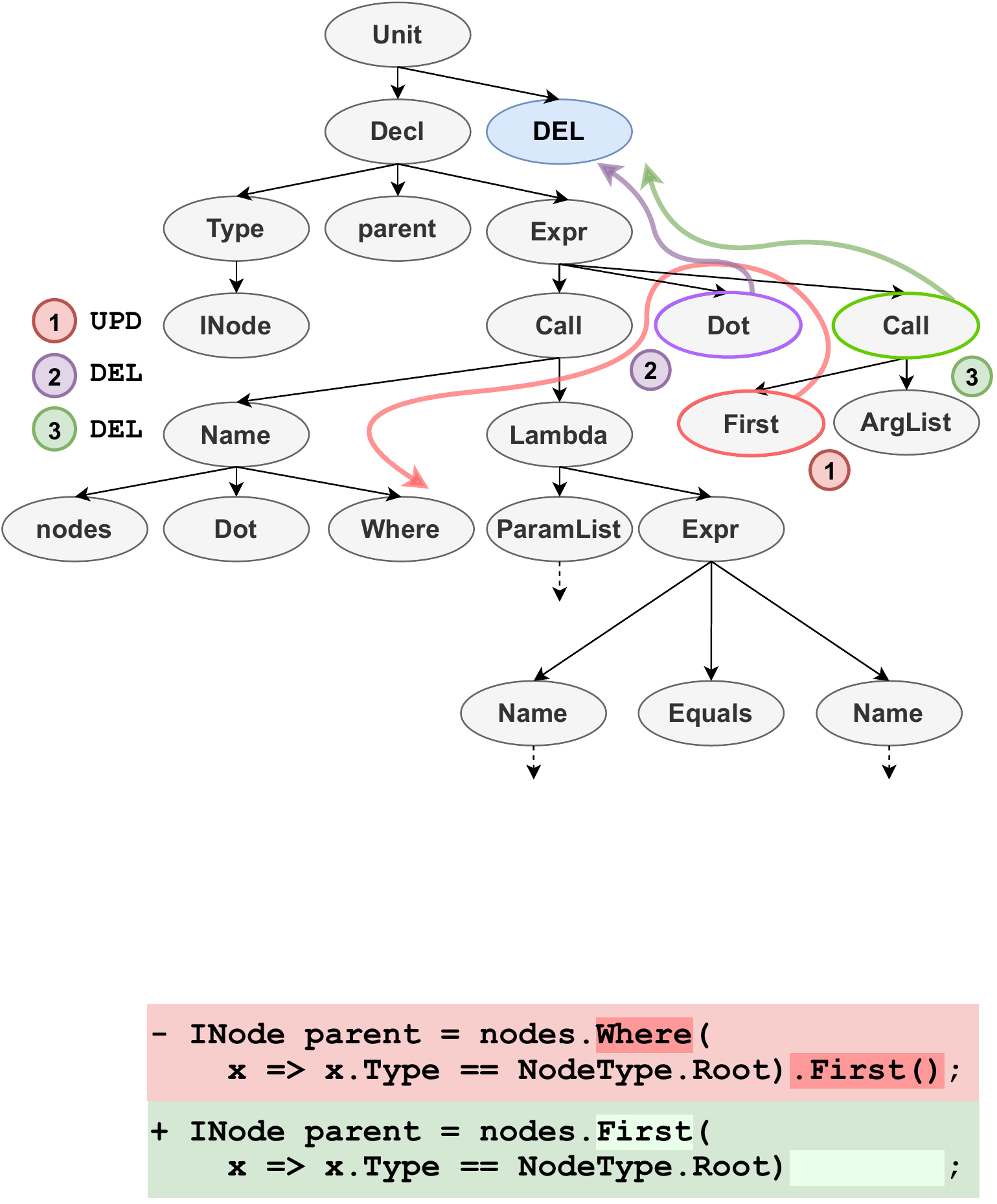}  
  \caption{}
  \label{Fi:gen_fig_b}
\end{subfigure}
\caption{
An example of two edits. These examples are different and the edits operate on different values. However, observing the structure of these edits reveals the similarity between them and allows a learning model to generalize better. This similarity is expressed as almost identical AST paths. For simplicity, only the program fragment that should be edited $\pbefore$ is shown, without the context $\cbefore$.
}
\label{Fi:gen_fig}
\end{figure}
Software development is an evolutionary process. Programs are being maintained, refactored, fixed, and updated on a continuous basis. Program \emph{edits} are therefore at the very core of software development. Poor edits can lead to bugs, security vulnerability, unreadable code, unexpected behavior, and more. The ability to suggest a good edit in code is therefore crucial.

We introduce the \ourtask{} task:
predict \emph{edit completions} based on a learned model that was trained on past edits. Given a code snippet that is partially edited, our goal is to predict an \emph{edit completion} that \emph{completes the edit for the rest of the snippet}. The edit completion is represented technically as a sequence of edit operations that we refer to as an \emph{edit script}.

\paragraph{Problem Definition}
Let $\pbefore$ be a given program fragment and $\cbefore$ be the surrounding context of $\pbefore$ before any edits were applied. Let $\Delta_\mathcal{C}$ denote the edits that were applied to $\cbefore$, and $\cafter=\Delta_\mathcal{C}\left(\cbefore\right)$ the resulting edited context.
The goal in our \ourtask{} task is to predict an edit function $\Delta_\mathcal{P}$, such that applying $\Delta_\mathcal{P}$ to $\pbefore$ results in the program fragment after the edit: $\Delta_\mathcal{P}\left(\pbefore\right)=\pafter$. 
Our underlying assumption is that the distribution of edits in $\pbefore$ can be inferred from the edits $\Delta_{\mathcal{C}}$ that occurred in its context.
We thus model the probability: $Pr\left(\Delta_\mathcal{P} \mid \Delta_\mathcal{C}\right)$. We present a new approach for representing and predicting $\Delta_\mathcal{P}$ in the \ourtask{} task, named \ctc{}: Contextual Code Changes. 

\paragraph{Motivating Examples}

Consider the \ourtask{} examples in \Cref{Fi:example_a} and \Cref{Fi:example_b}. These illustrate the significance of edits in the context $\cbefore$ and how they can help in suggesting a likely edit for $\pbefore$. In \Cref{Fi:example_a}, the edit in the context consists of changing the \scode{if} statement predicate, resulting in a \scode{null} check for the variable \scode{attack}. After the edit in the context, the value of \scode{attack} in $\pbefore$ cannot be \scode{null}. Therefore, the ternary statement that checks \scode{attack} for nullness in $\pbefore$ can be removed. Our model successfully predicted the needed edit $\Delta_\mathcal{P}$, which is applied to $\pbefore$ to yield $\pafter$. 

\Cref{Fi:example_b} shows another example, in which the edit in the context is a modification of a function signature. In $\cafter$, the return type was changed to \scode{FileCharacteristics}, and the output parameter \scode{fileCharacteristics} for the function was removed. $\pbefore$ consists of an assignment to the parameter \scode{fileCharacteristics}, and a return statement of \scode{true} value. The edit in the context implies a necessary edit in $\pbefore$, in which the assignment statement has to be removed (since \scode{fileCharacteristics} is no longer defined) and the return statement must include a variable of type \scode{FileCharacteristics}. Our model successfully predicted the correct edit for $\pbefore$. $\pafter$ consists of returning an object of type \scode{FileCharacteristics}.

\paragraph{Edit Completion vs. Code Completion}
It is important to note that \ourtask{} and code completion are completely different tasks. The goal of code completion is to predict missing fragments of a program, given a partial program as context. 
In contrast, the goal of \ourtask{} is to predict additional edits in a partial sequence of edit operations. That is, while code completion operates on code, \ourtask{} operates on \emph{code edits}.

\paragraph{Representing Code Edits}
The main design decision in learning code edits is \emph{how to represent the edit}, i.e., how to represent the difference between the code in its original form and its desired, altered, form.
Na\"ively, differencing programs can be performed by treating the code as text and using \emph{text-diff} algorithms for line differencing \citep{Hunt1975AnAF} or inline differencing \citep{10.1093/nar/24.14.2730}. In contrast, we model the \emph{difference between the abstract syntax trees} (ASTs) of the original and the edited code. This allows us to naturally use paths in the AST (AST paths) to model edits.

\paragraph{Our Approach}
We present a novel approach for \ourtask{}: predicting contextual code changes -- \ctc{}.
Code changes can be described as a sequence of edit operations, such as \emph{``move a node, along with its underlying subtree, to be a child of another node''} or \emph{``update the value of a node to be identical to the value of another node''}.
Such edit operations can be naturally represented as paths between the source node and the target node, along with the relationship between them and the edit command, i.e., ``move'' or ``update''. 
AST paths provide a natural way to express binary relationships between nodes (and thus subtrees) in the AST.
We use AST paths to represent $\Delta_\mathcal{C}$ -- edits that occurred in the context and transformed $\cbefore$ into $\cafter$, such that  $\Delta_\mathcal{C}\left(\cbefore\right)=\cafter$.
We also use AST paths to represent $\Delta_\mathcal{P}$ -- the edits that should be applied to $\pbefore$.
We thus model the probability $Pr\left(\Delta_\mathcal{P} \mid \Delta_\mathcal{C} \right)$, where both the input $\Delta_\mathcal{C}$ and the output $\Delta_\mathcal{P}$ are represented as AST paths.

Representing edits as paths allows a learning model to generalize well across different examples. Consider the two examples in \Cref{Fi:gen_fig}. 
In \Cref{Fi:gen_fig_a}, the edit modifies a series of LINQ calls,  converting \scode{Where(<predicate>).FirstOrDefault()} into  \scode{FirstOrDefault(<predicate>)}. 
The edit in \Cref{Fi:gen_fig_b} modifies \scode{Where(<predicate>).First()} into  \scode{First(<predicate>)}.
We elaborate on the representation of edits as paths in \Cref{Se:Overview} and \Cref{Se:EditAsPath}. For now, it suffices to note that there is a sequence of three edit operations in each of the figures (numbered \redone, \purpletwo, \greenthree).
Although the predicates are different and these edits operate on different values, the structure of the edits in \Cref{Fi:gen_fig_a} and \Cref{Fi:gen_fig_b} is identical. This similarity is expressed in the AST paths that represent these edits. For example, consider the identical structure of the path \redone{} in the two figures, where it operates on a different value in each figure (\scode{FirstOrDefault} and \scode{First}).

Our use of AST paths allows the model to generalize these edits, even though these edits are not identical and their predicates are different.

We apply a Pointer Network \citep{vinyals2015pointer} to point to paths in the AST of $\pbefore$ and create an edit operation sequence, i.e., an edit script.
While prior work used AST paths to \emph{read} programs and predict a label \cite{code2vec, alon2018codeseq}, we generate an edit script by \emph{predicting} AST paths, i.e., making AST paths \emph{the output} of our model.

\paragraph{Previous Approaches}
In related tasks, such as bug fixing and program repair, previous approaches have mostly represented code as a flat token stream
 \cite{DBLP:journals/corr/abs-1901-01808, DBLP:journals/corr/abs-1812-08693, vasic2018neural}; although this allows the use of NLP models out-of-the-box, such models do not leverage the rich syntax of programming languages. 

 \citet{yin2018learning} suggested a system that learns to represent an edit and uses its representation to apply the edit to another code snippet. Although it sounds similar, the task that \citet{yin2018learning} addressed and our task are dramatically different. \citet{yin2018learning} addressed the (easier) variant and assume that the edit that needs to be applied is \emph{given} as part of the input. This is done in the form of ``before'' and ``after'' versions of another code with \emph{the same edit applied}; their task is only to apply the given edit on a given code. 
 Thus, in the task of \citet{yin2018learning}, the assumption is that $\Delta_\mathcal{C} = \Delta_\mathcal{P}$.
In contrast, we do not assume that the edit $\Delta_\mathcal{P}$ is given; we condition on edits that occurred in the context ($\Delta_\mathcal{C}$), but these edits are different than the edits that need to be applied to $\pbefore$. Our model needs to predict the edit to $\pbefore$ itself, i.e., \emph{predict what needs to be edited and how}.
Other work did use syntax but did not represent the structure of the edit itself.
 \citet{Dinella2020HOPPITY:} proposed a model for detecting and fixing bugs using graph transformations, without considering context changes (i.e., $\Delta_{\mathcal{C}}=\emptyset$). Their method can predict \emph{unary} edit operations on the AST. In contrast, we predict \emph{binary} edit operations. Thus, our representation is much more expressive. %
 For example, consider the edit of moving a subtree. This edit can be represented as a \emph{single binary} operation; alternatively, this edit can require \emph{multiple unary} operations.

\paragraph{Modeling Code Likelihood vs. Modeling Edit Likelihood}
In general, there are two main approaches for learning to edit a given code snippet.
Assume that we wish to model the probability of a code snippet $\mathcal{Y}$ given another code snippet $\mathcal{X}$.
Much prior work \cite{DBLP:journals/corr/abs-1901-01808, 48350} followed the approach of generating $\mathcal{Y}$ directly.  Attempting to model $\mathcal{Y}$ given $\mathcal{X}$ modeled the probability $Pr\left(\mathcal{Y} \mid \mathcal{X}\right)$. 
This approach is straightforward, but requires modeling the likelihood of $\mathcal{Y}$, which is a problem that is more difficult than necessary.
In contrast, it can be much more effective to model \emph{the likelihood of the edit}, which transforms $\mathcal{X}$ into $\mathcal{Y}$, without modeling the likelihood of $\mathcal{Y}$ itself; hence, $Pr\left(\Delta_{\mathcal{X}\rightarrow\mathcal{Y}} \mid \mathcal{X}\right)$.
Our modeling of the \emph{edit} follows the latter approach: $Pr\left(\Delta_\mathcal{P} \mid \Delta_\mathcal{C} \right)$. %
In this work, we learn to predict the \emph{edit} ($\Delta_\mathcal{P}$) that transforms $\pbefore$ into $\pafter$, instead of predicting the entire program ($\pafter$). 
By applying $\Delta_\mathcal{P}$ to $\pbefore$, generating $\pafter$ is straightforward: $\Delta_\mathcal{P}\left(\pbefore\right)=\pafter$.
Learning to predict \emph{the edit} instead of learning to predict \emph{the edited code} makes our learning task much easier and provides much higher accuracy, as we show in \Cref{Se:Experiments}.

We show the effectiveness of \ctc{} on \ourtask{} on a new dataset, scraped from over 300,000 commits in GitHub. 

Our approach significantly outperforms textual and syntactic approaches that either model the code or model only the edit, and are driven by strong neural models.

\paragraph{Contributions}
The main contributions of this paper are: 
\begin{itemize}
    \item We introduce the \ourtask{} task: given a program $\pbefore$ and edits that occurred in its context, predict the likely edits that should be applied to $\pbefore$.
    \item \ctc{} -- a novel approach for representing and predicting contextual edits in code. This is the first approach that represents structural edits directly. 
    \item Our technique directly captures the relationships between subtrees that are changed in an edit using \emph{paths} in the AST. The output of our technique is an edit script that is executed to edit the program fragment $\pbefore$. %
    \item A prototype implementation of our approach, called C$^3$PO, for Contextual Code Changes via Path Operations. 
    C$^3$PO is implemented using a strong neural model that predicts the likely edit by pointing to an AST path that reflects that edit. 
    \item A new \ourtask{} dataset of source code edits and their surrounding context edits, scraped from over 300,000 commits in GitHub. 
    \item An extensive empirical evaluation that compares our approach to a variety of representation and modeling approaches, driven by strong models such as LSTMs, Transformers, and neural CRFs. Our evaluation shows that our model achieves over $28\%$ relative gain over state-of-the-art strong sequential models, and over $2\times$ higher accuracy than syntactic models that do not model edits directly.
    \item A thorough ablation study that examines the contribution of syntactic and textual representations in different components of our model.
\end{itemize}

\section{Overview}\label{Se:Overview}

\begin{figure}[t]
\begin{subfigure}[b]{1\textwidth}
  \centering
  \includegraphics[width=1\linewidth]{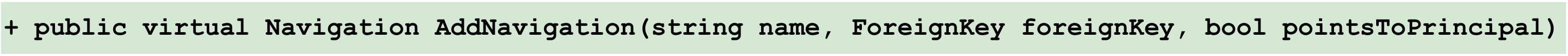}  
  \caption{}
  \label{Fi:ctx_changes}
\end{subfigure}

\begin{subfigure}[b]{.49\textwidth}
  \centering
  \includegraphics[width=1\linewidth]{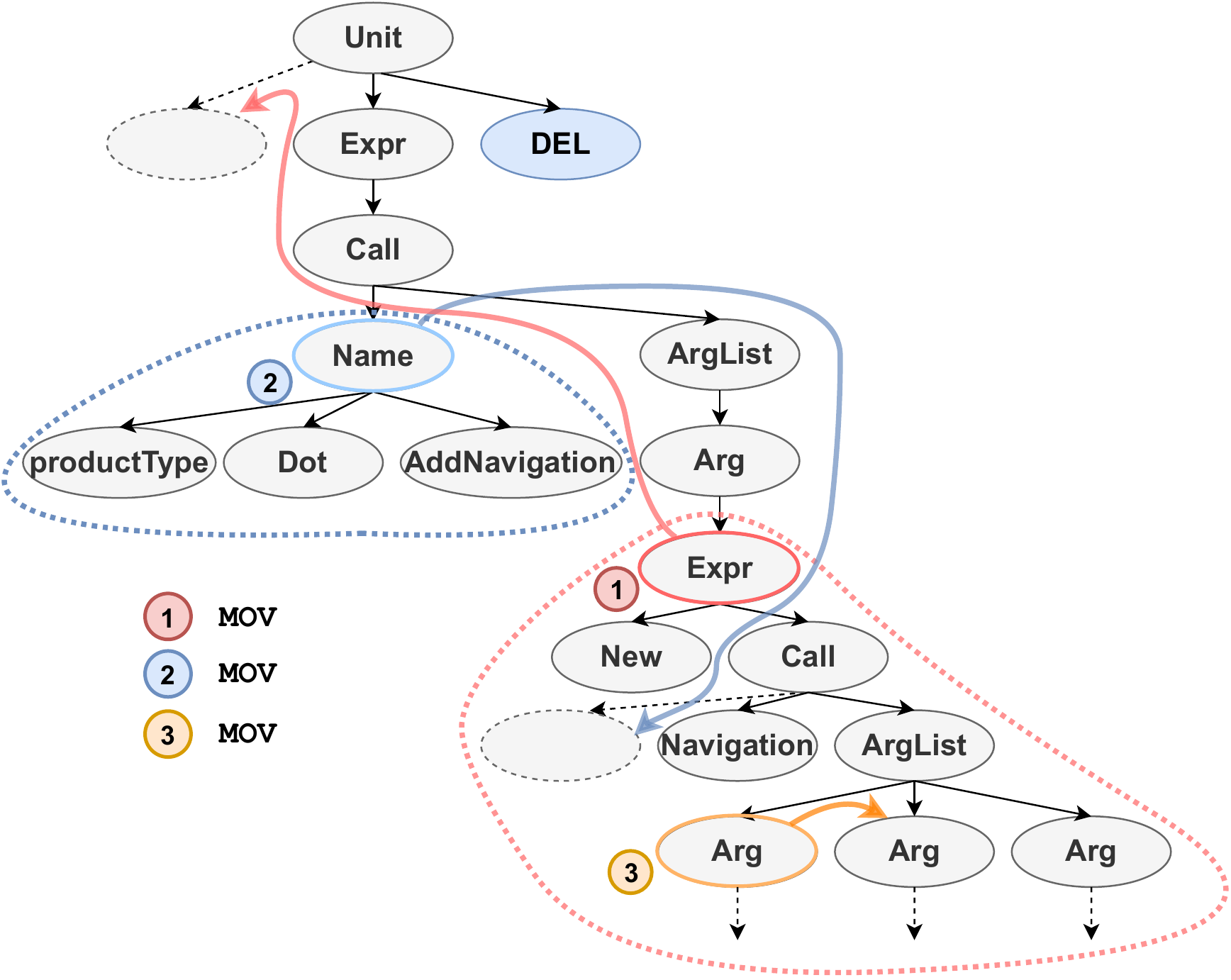}  
  \caption{}
  \label{Fi:first_3op}
\end{subfigure}
\begin{subfigure}[b]{.49\textwidth}
  \centering
  \includegraphics[width=1\linewidth]{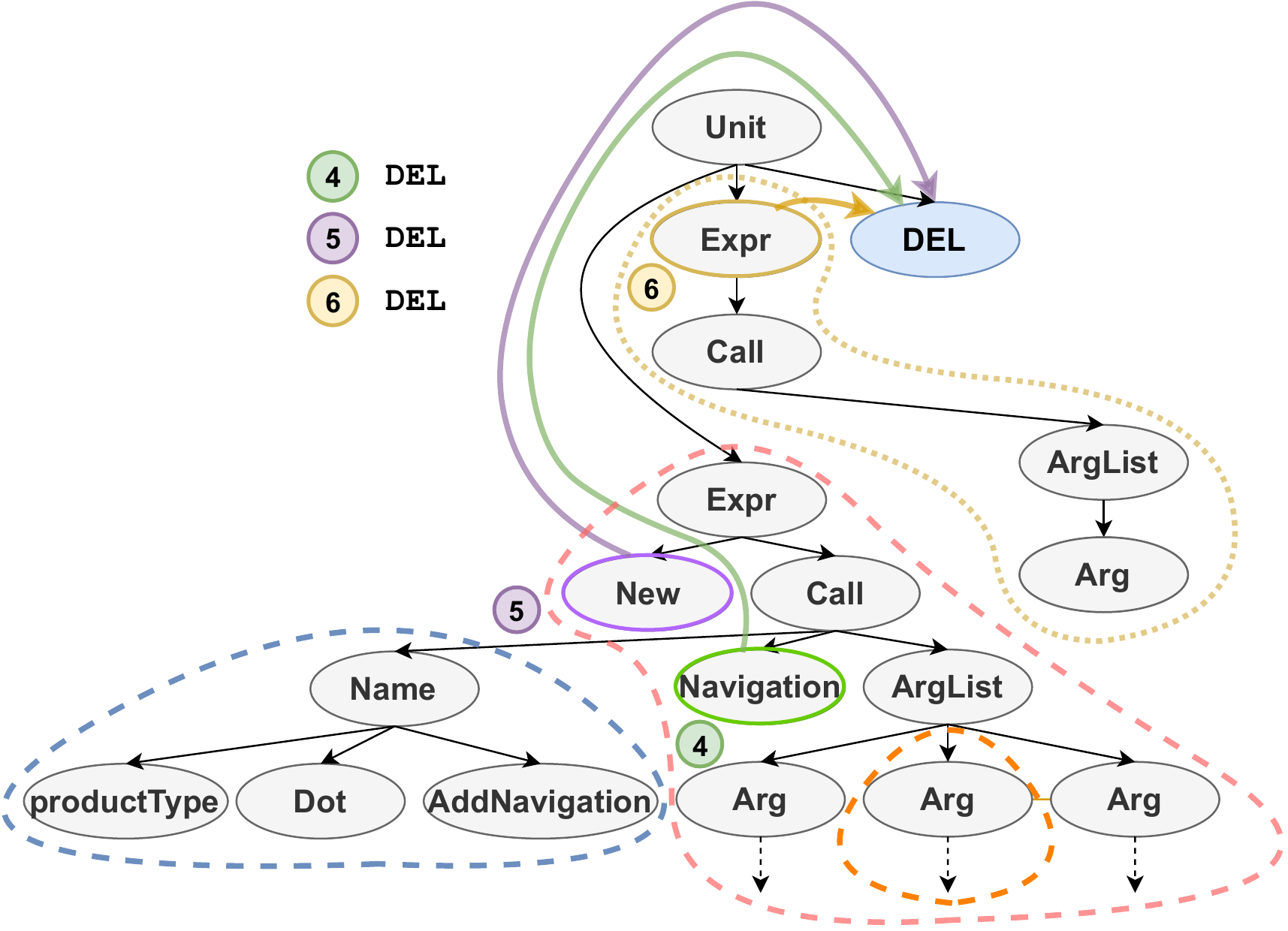}  
  \caption{}
  \label{Fi:last_3op}
\end{subfigure}

\begin{subfigure}[b]{.49\textwidth}
  \centering
  \includegraphics[width=1\linewidth]{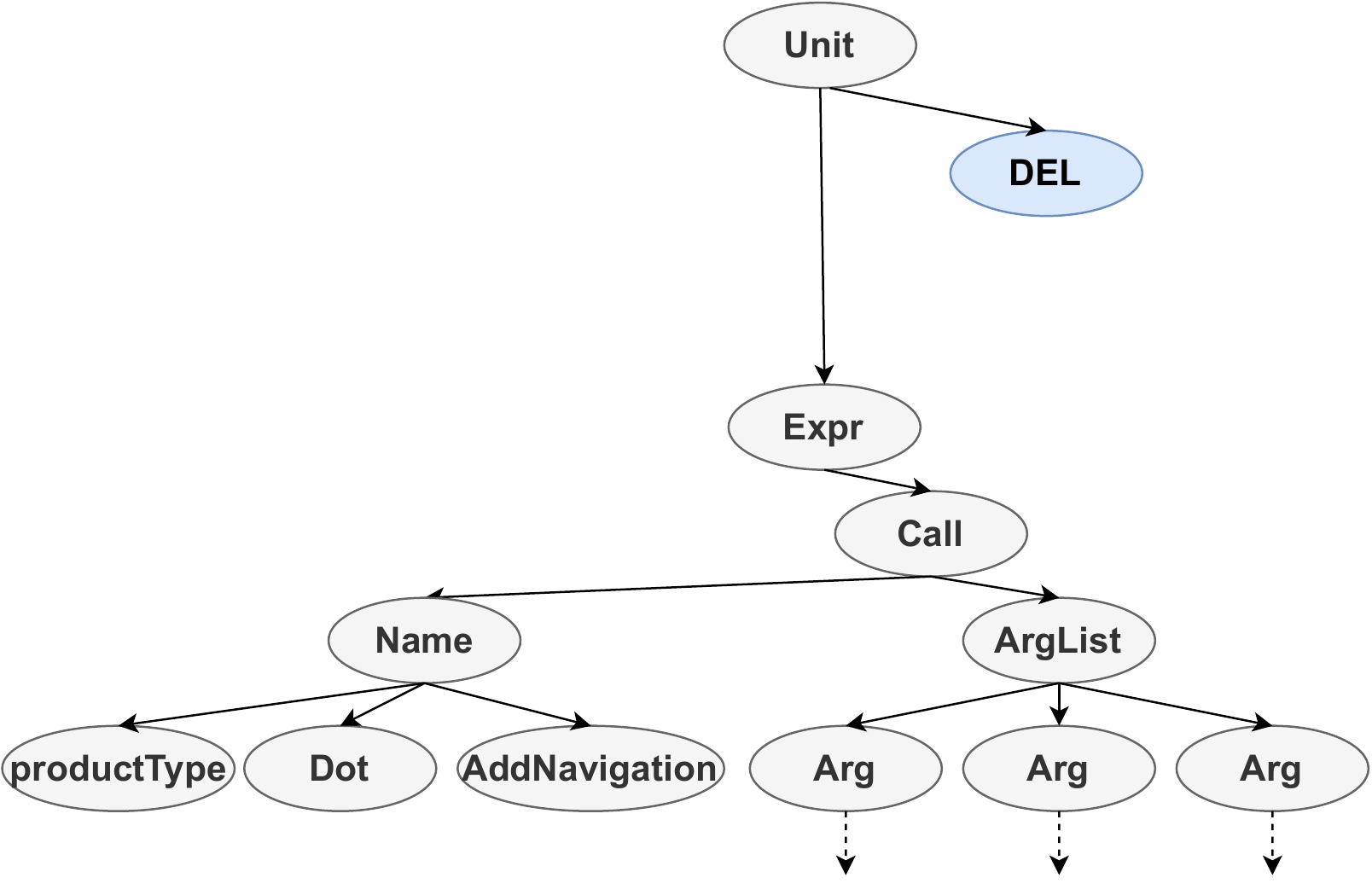}  
  \caption{}
  \label{Fi:final_tree}
\end{subfigure}
\begin{subfigure}[b]{0.49\textwidth}
  \centering
  \raisebox{8mm}{
  \includegraphics[width=0.83\linewidth]{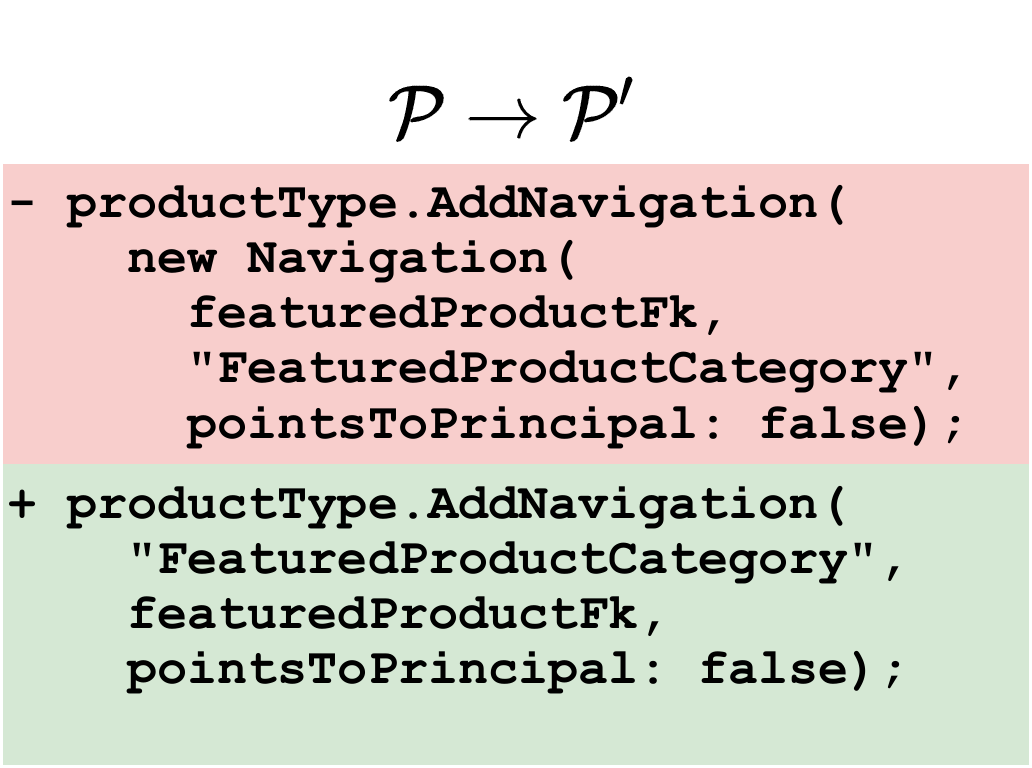}  
  }
  \caption{}
  \label{Fi:overview_line_diff}
\end{subfigure}
\caption{ 
An \ourtask{} example from our test set. 
\Cref{Fi:ctx_changes} shows the edit that transforms $\cbefore$ into $\cafter$ -- overloading the function \scode{AddNavigation}.
\Cref{Fi:overview_line_diff} shows $\pbefore$ and $\pafter$ as code in red and green, respectively. 
\Cref{Fi:first_3op} depicts the partial AST and the first three edit operations of the edit. \Cref{Fi:last_3op} shows the AST after applying the first three operations, and shows the next three operations as AST paths. 
\Cref{Fi:final_tree} illustrates the AST after performing all operations, resulting in an AST that corresponds to $\pafter$. 
Every edit operation is represented by an AST path having the same color and number as the edit command. Dotted contours represent subtrees that will be affected by applying these operations. 
Already-affected subtrees are surrounded by dashed contours. 
}
\label{Fi:overview}
\end{figure}
In this section, we demonstrate our approach using a simple \ourtask{} example. 
The main idea is to represent all valid edit operations in $\pbefore$ as AST paths, %
and predict a sequence of these paths.
Since every path is associated with an edit operation, by pointing to a sequence of paths, we, in fact, predict an edit script.

\subsection{Motivating Example}
\paragraph{High-level Overview}
Consider the edit that occurred in the context of \Cref{Fi:ctx_changes} -- insertion of a new definition of the method \scode{AddNavigation}, which overloads previous definitions. After applying this edit, it is possible to use this new signature when calling \scode{AddNavigation}.
Consider the original code snippet $\pbefore$ at the top of \Cref{Fi:overview_line_diff}. 
The edit in the context allows us to simplify the call to \scode{AddNavigation}  using the new signature, as shown in the ``edited'' code snippet $\pafter$ at the bottom of \Cref{Fi:overview_line_diff}. Consider the partial AST of $\pbefore$ in \Cref{Fi:first_3op}.
The desired edit can be described as an edit script consisting of six edit operations to the AST of $\pbefore$. Consider the first operation: \textbf{\textcolor{Maroon}{\raisebox{.5pt}{\textcircled{\raisebox{-.9pt} {1}}}}} \scode{\textbf{MOV}}.
The meaning of this operation is to move the node \scode{Expr} with its subtree to be the leftmost child of the node \scode{Unit}. This edit operation can be represented by the \textbf{\textcolor{Maroon}{red\raisebox{.5pt}{ \textcircled{\raisebox{-.9pt} {1}}}}} path: \scode{Expr} $\rightarrow$ \scode{Arg} $\rightarrow$ \scode{ArgList} $\rightarrow$ \scode{Call} $\rightarrow$ \scode{Expr} $\rightarrow$ \scode{Unit}.
Note how this path directly captures the syntactic relationship between the node \scode{Expr} and the node \scode{Unit}, allowing our model to predict 
a \textbf{\scode{MOV}} operation
as part of the edit script. 

In \Cref{Fi:last_3op} we can see the result of applying the following first three operations: 
 \textbf{\textcolor{Maroon}{\raisebox{.5pt}{\textcircled{\raisebox{-.9pt} {1}}}}} \textbf{\scode{MOV}}, \textbf{\textcolor{NavyBlue}{\raisebox{.5pt}{\textcircled{\raisebox{-.9pt} {2}}}}} \textbf{\scode{MOV}}, \textbf{\textcolor{YellowOrange}{\raisebox{.5pt}{\textcircled{\raisebox{-.9pt} {3}}}}} \textbf{\scode{MOV}}, moving subtrees to new locations in the tree. The last three commands are \textbf{\scode{DEL}} operations, expressing deletion of a node and its underlying subtree. These operations can be represented using paths as well. For instance, \textbf{\textcolor{LimeGreen}{\raisebox{.5pt}{\textcircled{\raisebox{-.9pt} {4}}}}} \textbf{\scode{DEL}} is represented by the 
\textbf{\textcolor{LimeGreen}{green\raisebox{.5pt}{ \textcircled{\raisebox{-.9pt} {4}}}}}
path: \scode{Navigation} $\rightarrow$ \scode{Call} $\rightarrow$ \scode{Expr} $\rightarrow$ \scode{Unit} $\rightarrow$ \scode{DEL}, where \scode{DEL} is an artificial node that we add as a child of the AST's root. In \Cref{Fi:final_tree} we can see the AST after applying all six operations. %
After executing all six operations, our model produces $\pafter$, shown in \Cref{Fi:overview_line_diff}.

\paragraph{Path Extraction}
To inform the model about the available edits it can use for prediction,
we parse the AST of $\pbefore$ to extract all AST paths that represent valid edits. 
Every path can represent different edit ``commands'' that use the same path.
For example, consider the \textbf{\textcolor{NavyBlue}{blue\raisebox{.5pt}{ \textcircled{\raisebox{-.9pt} {2}}}}} path in \Cref{Fi:first_3op}: \scode{Name} $\rightarrow$ \scode{Call} $\rightarrow$ \scode{ArgList} $\rightarrow$ \scode{Arg} $\rightarrow$ \scode{Expr} $\rightarrow$ \scode{Call}. %
This path can represent a move operation -- \textbf{\scode{MOV}}, i.e., moving the node \scode{Name} with its subtree, to be the leftmost child of \scode{Call}; alternatively, this path can represent an insertion operation -- \textbf{\scode{INS}}, i.e., copy \scode{Name} with its subtree, and insert it as the leftmost child of \scode{Call}. 
To distinguish between different edit operations that are represented using the same AST path, each path is encoded as a vector once, and projected into three vectors using different learned functions. Each resulting vector corresponds to a different kind of edit operation. For example, the
\textbf{\textcolor{YellowOrange}{orange\raisebox{.5pt}{ \textcircled{\raisebox{-.9pt} {3}}}}} path in \Cref{Fi:first_3op} can represent either ``move'' (\textbf{\scode{MOV}}), ``update'' (\textbf{\scode{UPD}}) or ``insert'' (\textbf{\scode{INS}}) operations. In this case, this path was projected using the learned function that represents ``move''.

\paragraph{Edit Script Prediction}
We predict one edit operation at each step by \emph{pointing} at a path and its associated operation from among the valid edit operations. This results in an \emph{edit script}. For example, in \Cref{Fi:overview}, our model finds that the \textbf{\textcolor{Maroon}{red\raisebox{.5pt}{ \textcircled{\raisebox{-.9pt} {1}}}}} path with \textbf{\scode{MOV}} is most likely to be the first operation. Then, given this edit, our model finds that the  \textbf{\textcolor{NavyBlue}{blue\raisebox{.5pt}{ \textcircled{\raisebox{-.9pt} {2}}}}} path with \textbf{\scode{MOV}} is most likely to be the next operation, and so on, until we predict a special ``end of sequence'' (\scode{EOS}) symbol.

\paragraph{Modeling Code Likelihood vs. Modeling Edit Likelihood}
Modeling edits using AST paths provides an effective way to model \emph{only the difference} between $\pbefore$ and $\pafter$. For example, consider the \textbf{\textcolor{Maroon}{red\raisebox{.5pt}{ \textcircled{\raisebox{-.9pt} {1}}}}} path that moves the subtree rooted at \scode{Expr} from its original place to be the first child of \scode{Unit}. 
To predict this edit, our model only needs to select the \textbf{\textcolor{Maroon}{red\raisebox{.5pt}{ \textcircled{\raisebox{-.9pt} {1}}}}} path out of the other available operations.
In contrast, a model that attempts to generate $\pafter$ entirely \cite{DBLP:journals/corr/abs-1901-01808}, would need to generate the entire subtree from scratch in the new location.

\paragraph{Pairwise Edit Operations}
Most edit operations, such as ``move'' and ``update'', can be described as \emph{pairwise} operations, having the ``source'' and the ``target'' locations as their two arguments. AST paths provide a natural way to represent pairwise relations, originating from the ``source'' location, and reaching the ``target'' location through the shortest path between them in the tree. In contrast, prior work that used only \emph{unary} edit operations such as HOPPITY \cite{Dinella2020HOPPITY:} are limited to inserting each node individually, and thus use \emph{multiple} edit commands to express the  \textbf{\textcolor{Maroon}{\raisebox{.5pt}{\textcircled{\raisebox{-.9pt} {1}}}}} \textbf{\scode{MOV}} operation. Our model represents this edit operation as a single AST path -- the \textbf{\textcolor{Maroon}{red\raisebox{.5pt}{ \textcircled{\raisebox{-.9pt} {1}}}}} path.

\paragraph{Key aspects} The example in \Cref{Fi:overview} demonstrates several key aspects of our method:
\begin{itemize}
    \item Edits applied to the context of $\pbefore$ can provide useful information for the required edit to $\pbefore$.
    \item Pairwise edit operations can be naturally represented as AST paths.
    \item A neural model, trained on these paths, can generalize well to other programs, thanks to the direct modeling of code edits as paths.
    \item By \emph{pointing} at the available edit operations, the task that the model addresses becomes \emph{choosing} the most likely edit, rather than generating $\pafter$ from scratch, and thus significantly eases the learning task.
\end{itemize}

\section{Background}
In this section, we provide the necessary background. First we define abstract syntax trees (ASTs) and AST paths  in \Cref{Se:ASTPaths}. In \Cref{Se:ASTDiff}, we use these definitions to describe how to represent code edits using AST paths and perform AST differencing. Finally, in \Cref{Se:AttnPtr} and \Cref{Se:Ptr}, we describe the concept of \emph{attention} and \emph{pointer networks}, which are crucial components in our neural architecture (described in \Cref{Se:Arch}). 

\subsection{AST Paths}\label{Se:ASTPaths}
Given a programming language $\mathcal{L}$ and its grammar, we use $V$ to denote the set of \textit{nonterminals}, and $T$ to denote the set of \textit{terminals} in the grammar. The Abstract Syntax Tree (AST) of a program can be constructed in the standard manner, defined as follows: 
\begin{definition}
(Abstract Syntax Tree) Given a program $\mathcal{P}$ written in a programming language $\mathcal{L}$, its Abstract Syntax Tree $\mathcal{A}$ is the tuple $(A,B,r,X,\delta, \phi)$, where $A$ is the set of non-leaf nodes, such that each $n\in A$ is of type that belongs to $V$; 
$B$ is the set of leaves such that each $n\in B$ is of type that belongs to $T$; $r \in A$ is the root of the tree;
$X$ is a set of values taken from $\mathcal{P}$;
$\delta$ is a function $\delta: A\rightarrow (A\cup B)^*$ that maps \textit{nonterminals} nodes to their children; 
$\phi$ is a mapping $\phi: B\rightarrow X$ that maps a \textit{terminal} node to a value.
\end{definition}

An AST path is simply a sequence of nodes in the AST, formally:
\begin{definition}
(AST Path) Given an AST $\mathcal{A}=(A,B,r,X,\delta, \phi)$, an AST path is a sequence of nodes $p=n_1,n_2,...,n_k$, where $n_i\in A \cup B$, such that for every consecutive pair of nodes $n_i$ and $n_{i+1}$, either $n_i\in\delta(n_{i+1})$ or $n_{i+1}\in\delta(n_{i})$. We follow \citet{Alon_2018} and associate each node's \emph{child index} with its type.
\end{definition}

For example, consider the \textbf{\textcolor{NavyBlue}{blue\raisebox{.5pt} { \textcircled{\raisebox{-.9pt} {2}}}}} path in Figure \ref{Fi:first_3op}. 
The path starts in the node \scode{Name}, goes up to its parent node \scode{Call}, then goes down to its right-most child \scode{ArgList}, an so on. %

AST paths are a natural way to describe relationships between nodes in the AST, and can serve as a general representation of relationships between elements in programs. For example, \citet{Alon_2018,code2vec} used paths between leaves in the AST as a way to create an aggregated representation of the AST. 

In this work, we use AST paths to model relationships between arbitrary nodes in the tree (both terminals and nonterminals) to model the effect of edit operations.

\begin{figure}
\begin{subfigure}[b]{.19\textwidth}
  \centering
  \includegraphics[scale=0.6]{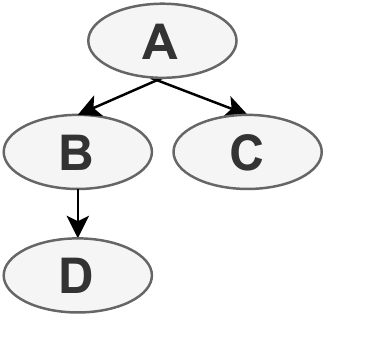}  
  \caption{}
  \label{Fi:ast_diff}
\end{subfigure}
\hfill
\begin{subfigure}[b]{.19\textwidth}
  \centering
  \includegraphics[scale=0.6]{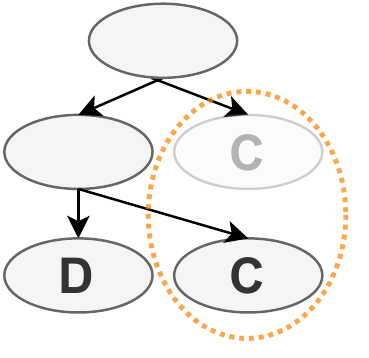}  
  \caption{\textbf{\scode{MOV}}}
  \label{Fi:ast_diff_mov}
\end{subfigure}
\hfill
\begin{subfigure}[b]{.19\textwidth}
  \centering
  \includegraphics[scale=0.6]{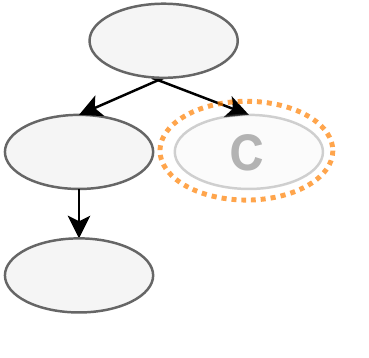}  
  \caption{\textbf{\scode{DEL}}}
  \label{Fi:ast_diff_del}
\end{subfigure}
\begin{subfigure}[b]{.19\textwidth}
  \centering
  \includegraphics[scale=0.6]{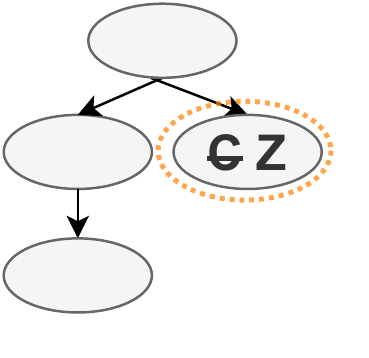}  
  \caption{\textbf{\scode{UPD}}}
  \label{Fi:ast_diff_upd}
\end{subfigure}
\begin{subfigure}[b]{.19\textwidth}
  \centering
  \includegraphics[scale=0.6]{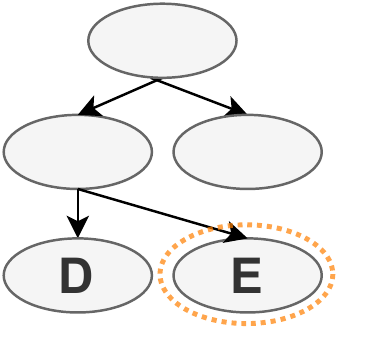}  
  \caption{\textbf{\scode{INS}}}
  \label{Fi:ast_diff_ins}
\end{subfigure}
\caption{
Example of AST edit operations. \Cref{Fi:ast_diff} depict the AST before the change. \Cref{Fi:ast_diff_mov} shows the result of \textbf{\scode{MOV}} operation -- moving \scode{C} to be the right sibling of \scode{D}. \Cref{Fi:ast_diff_del} shows the result of \textbf{\scode{DEL}} -- removing \scode{C}. \Cref{Fi:ast_diff_upd} shows the result of \textbf{\scode{UPD}} -- updating \scode{C} to \scode{Z}. \Cref{Fi:ast_diff_ins} shows the result of \textbf{\scode{INS}} -- inserting \scode{E} to be the right sibling of \scode{D}.
}
\label{Fi:ast_diff_ops}
\end{figure}

\subsection{AST Differencing}\label{Se:ASTDiff}
An edit in a program can be represented as a sequence of operations on its AST. 
To compute the difference between two programs, we 
compute the difference between the two ASTs using algorithms such as GumTree \citep{DBLP:conf/kbse/FalleriMBMM14}.
Given two programs $\pbefore$ and $\pafter$, along with their ASTs $\mathcal{A}$ and $\mathcal{A}'$, GumTree outputs an \emph{edit script}, consisting of instructions to  change $\mathcal{A}$ so it becomes $\mathcal{A}'$. Each operation in the script is either \textbf{\scode{MOV}}, \textbf{\scode{DEL}}, \textbf{\scode{UPD}}, or \textbf{\scode{INS}} and operates on one or two nodes. 
The command \textbf{\scode{MOV}} $n_s, n_t$ stands for moving a subtree inside the AST. This operation takes the source node $n_s$ to be moved and the target $n_t$ node, which will be the left sibling of $n_s$ after the move.
The command \textbf{\scode{DEL}} $n_s$ stands for removing the node $n_s$ from the tree. 
We use the command \textbf{\scode{UPD}} $v, n_t$, to update the value of the node $n_t$ to become $v$. 
Lastly, to represent insertion, we use  \textbf{\scode{INS}} $n_s, n_t$, where $n_s$ is the root of a subtree to be inserted and $n_t$ is the target node that will be the left sibling of $n_s$ after the insertion. 

\Cref{Fi:ast_diff_ops} demonstrates all operations: \Cref{Fi:ast_diff} illustrates the AST before the edits;
\Cref{Fi:ast_diff_mov} shows the result of \textbf{\scode{MOV}} \scode{C}, \scode{D};
\Cref{Fi:ast_diff_del} depicts the command \textbf{\scode{DEL}} \scode{C};
\Cref{Fi:ast_diff_upd} shows the update of \scode{C} to the value \scode{Z}, i.e., \textbf{\scode{UPD}} \scode{Z}, \scode{C}; \Cref{Fi:ast_diff_ins} illustrates the command \textbf{\scode{INS}} \scode{E}, \scode{D} -- the insertion of node \scode{E} as a right sibling of \scode{D}.

In general, AST differencing algorithms consist of two steps. The first step maps nodes from $\mathcal{A}$ to $\mathcal{A}'$, where each node belongs to a single mapping at most and mapped nodes share the same type. The second step uses the mapping and aims to produce a short edit script. The GumTree algorithm focuses on the first step of mapping, since there are known quadratic optimal algorithms \citep{10.1145/235968.233366} for the second step. 

GumTree \cite{DBLP:conf/kbse/FalleriMBMM14} breaks the mapping stage into three steps. The first step is a top-down algorithm that finds isomorphic subtrees across $\mathcal{A}$ and $\mathcal{A}'$. The roots of these subtrees are called \emph{anchors mapping}. 
The second step is a bottom-up algorithm that looks for \emph{containers mapping}; these are node pairs among $\mathcal{A}$ and $\mathcal{A}'$, such that their descendants share common \emph{anchors}. Finally, the last step looks for additional mappings between the descendants of the \emph{containers mapping} pairs.

Applying the GumTree algorithm for the mapping stage and using known techniques to produce the edit script results in an end-to-end efficient algorithm. The complexity of this algorithm is $O(n^2)$ in the worst case, where $n$ is the number of nodes in the larger among $\mathcal{A}$ and $\mathcal{A}'$, i.e.,  $n=max(|\mathcal{A}|,|\mathcal{A}'|)$.

\subsection{Attention}\label{Se:AttnPtr}
An attention mechanism computes a learned weighted average of some input vectors, given another input \emph{query} vector.
Usually, attention is used by a neural model to align elements from different modalities. For example, in neural machine translation (NMT) \citep{bahdanau2014neural}, attention allows the model to ``focus'' on different words from the source language while predicting every word in the target language, by computing a different weighted average at every step.
This ability has shown a significant improvement across various tasks such as translation \citep{DBLP:journals/corr/LuongPM15,bahdanau2014neural, NIPS2017_7181}, speech recognition \cite{chan2016listen}, and code summarization and captioning \cite{alon2018codeseq}.

Formally, given a set of $k$ vectors $Z=\mathbf{z_1},\mathbf{z_2},..,\mathbf{z_k}\in \mathcal{R}^{d}$ (usually, an encoding of the input of the model) and a query vector $\mathbf{q}\in \mathcal{R}^{d}$ (usually, the hidden state of a decoder at a certain time step $t$), attention performs the following computation. The first step computes a ``score'' for each input vector $\mathbf{z}_i$. For example, \citet{DBLP:journals/corr/LuongPM15} use a learned matrix $W_a \in \mathcal{R}^{d\times d}$ to compute the score $s_i$ of the vector $\mathbf{z}_i$: %
\begin{equation}\label{Eq:att_1}
    s_i = \mathbf{z}_i \cdot W_a \cdot \mathbf{q}^{\top}
\end{equation}
Next, all scores are normalized into a pseudo-probability using the softmax function:
\begin{equation} \label{Eq:att_2}
    \alpha_i =  \frac{e^{s_i}}{\sum_{j=1}^{k} e^{s_j}}
\end{equation}
where every normalized score is between zero and one $\alpha_i \in [0,1]$, and their sum is one: $\sum{\alpha_i}=1$.
Then, a \emph{context vector} is computed as a weighted average of the inputs $\mathbf{z_1},\mathbf{z_2},..,\mathbf{z_k}$, such that the weights are the computed weights $\alpha$:
\begin{equation*}
    \mathbf{c} =  \sum_{i}^{k} \alpha_i \cdot \mathbf{z_i}
\end{equation*}

This dynamic weighted average can be computed iteratively at different prediction time steps $t$, producing different attention scores $\alpha_t$ and thus a different context vector $\mathbf{c_t}$. This offers a decoder the ability to focus on different elements in the encoded inputs at each prediction step. 

\subsection{Pointer Networks}\label{Se:Ptr}
A \emph{pointer network} \citep{vinyals2015pointer} is a variant of the seq2seq paradigm \cite{DBLP:journals/corr/SutskeverVL14}, where the output sequence is a series of \emph{pointers} to the encoded inputs, rather than a sequence from a separate vocabulary of symbols. %
This mechanism is especially useful when the output sequence is composed \emph{only} of elements from the input, possibly permutated and repeated.
For example, the problem of \emph{sorting} a sequence of numbers can be naturally addressed using pointer networks: the input for the model can be the unsorted sequence of numbers, and the output is the sorted sequence, where every output prediction is a \emph{pointer} to an element in the input sequence.

Pointing can be performed in a manner similar to attention:
at each decoding step, \Cref{Eq:att_1} and \Cref{Eq:att_2} compute input scores, similar to attention. Then, the resulting normalized scores $\alpha_i$ can be used for classification over the encoded inputs, as the output probability of the model. 

Pointer networks and attention share almost the same implementation, but they are different in principle. 
Attention computes a dynamic average $\mathbf{c}_t$ at each decoding iteration. Then, $\mathbf{c}_t$ is used in the prediction of this time step, among a different closed set of possible classes. For example, the possible classes can be the words in the target language. 
In pointer networks, on the other hand, the possible classes at each decoding step are the elements \emph{in the input sequence itself}. %

Another difference is that in pointer networks there is a label associated with each ``pointing'' step. Each ``pointing'' distribution $\alpha$ is directly supervised by computing a cross-entropy loss with a reference label. 
In other words, each pointing can be measured for its correctness, and the mostly-pointed input is either correct or incorrect.
In contrast, attention is not directly supervised; the model's attention distribution $\alpha$ is internal to the model. The attention distribution $\alpha$ is usually neither ``correct'' nor ``incorrect'', because the attention is used for a follow-up prediction.

\section{Representing Edits with AST Paths}\label{Se:EditAsPath}

In the \ourtask{} task that we consider in this work, the input contains multiple edit operations that occurred in the context, and the output is a series of edit operations that should be performed.
The main challenge is \emph{how to represent  edits in a learning model}? We look for a representation that is \emph{expressive} and \emph{generalizable}. The representation should be \emph{expressive}, such that different edits are reflected differently; this would allow a model to consider the difference between examples.
However, just representing every edit uniquely is not enough, because the representation should also be \emph{generalizable}, such that similar edits would be reflected similarly. This would allow a model to generalize better, even if the edit that should be predicted at test time does not look exactly like an edit that was observed at training time.

Representing edits using AST paths provides an expressive and generalizable solution. An edit operation, such as ``move'', can be represented as the path in the AST from the subtree that should be moved, up to its new destination. This path includes the \emph{syntactic relation} between the source and the target of the move. Different move operations would result in different paths and thus, this representation is expressive. Similar moves will result in similar paths and thus, this representation is generalizable.
In this section, we explain how AST paths can naturally represent such edit operations.

We represent edit operations as follows:
\begin{enumerate}
    \item The \textbf{\scode{MOV}} (move) operation has two arguments: the first is the source node -- the root of the subtree to be moved, and the second is the target node. The meaning of ``\textbf{\scode{MOV}} $n_s, n_t$'' is that node $n_s$ moves to be the right sibling of node $n_t$. To support moving a node to be the leftmost child, we augment the AST with \scode{Placeholder} nodes, that are always present as the leftmost child nodes of all \emph{nonterminal} nodes. 

    \item The \textbf{\scode{UPD}} (update) operation has two arguments: 
    the first argument is a node with a source value, and the second argument is a node whose value needs to be updated. For instance, if the value of node $n_t$ needs to be updated to \scode{x}, and the value of node $n_s$ is \scode{x}, we denote this by: ``\textbf{\scode{UPD}} $n_s, n_t$''. 
    \item The \textbf{\scode{INS}} (insert) operation has two arguments: the first argument is the subtree to be copied, and the second is the target node. The operation ``\textbf{\scode{INS}} $n_s, n_t$'' means that the subtree rooted at $n_s$ should be copied and inserted as the right sibling of $n_t$. If $n_s$ should be inserted as a leftmost child, the target node will be the appropriate \scode{Placeholder} node.
    \item The \textbf{\scode{DEL}} (delete) operation has one argument, which is a subtree to be deleted. We represent \textbf{\scode{DEL}} as a path that originates from the root of the subtree to be deleted, into a special \scode{DEL} target node that we artificially add as a child of the AST's root. So in practice, we represent ``\textbf{\scode{DEL}} $n_s$'' as ``\textbf{\scode{MOV}} $n_s, n_{\scode{DEL}}$'' where $n_{\scode{DEL}}$ is the \scode{DEL} node.
\end{enumerate}
Since all four operations can be represented using two nodes $n_s$ and $n_t$ from the AST of $\pbefore$, 
the AST path from $n_s$ to $n_t$ is a natural way to represent an edit operation. \Cref{Fi:editaspath} demonstrates a \textbf{\scode{MOV}} operation and its associated path representation. \Cref{Fi:mov_exmp_a} depicts the path  \scode{Arg$_1$} $\rightarrow$ \scode{ArgList} $\rightarrow$ \scode{Arg$_3$}, which can be associated with \textbf{\scode{MOV}} and represent the operation \textbf{\scode{MOV}} \scode{Arg$_1$, Arg$_3$}, i.e., moving the first argument to be the last. 
\Cref{Fi:mov_exmp_b} shows the AST after the movement.

\newcommand{\figfivescale}{0.5}
\begin{figure}
\begin{subfigure}[b]{.49\textwidth}
  \centering
  \includegraphics[scale=\figfivescale]{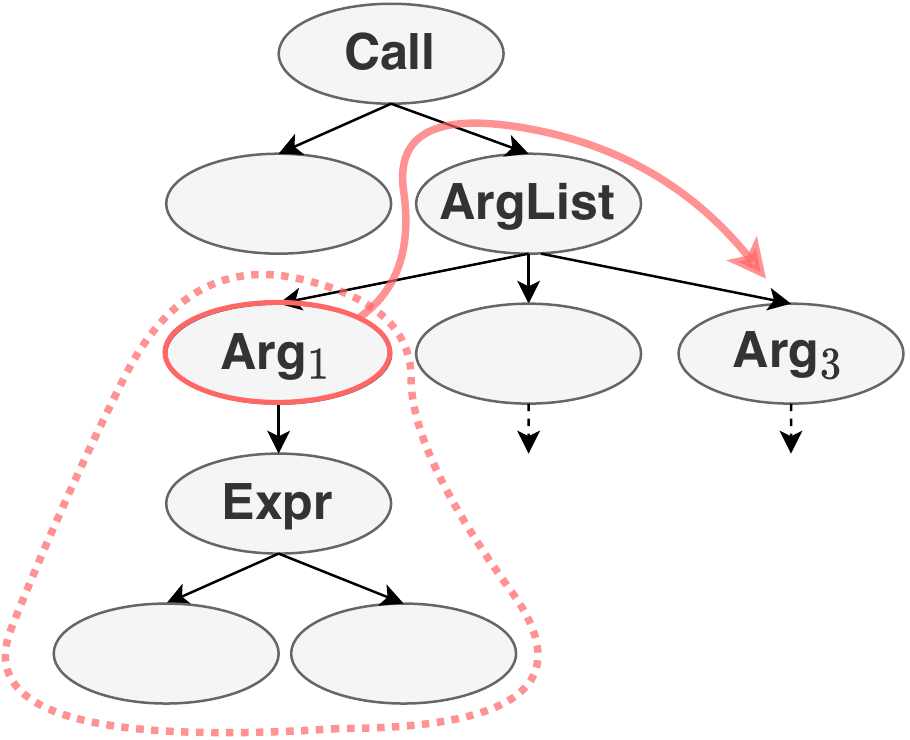}  
  \caption{}
  \label{Fi:mov_exmp_a}
\end{subfigure}
\hfill
\begin{subfigure}[b]{.49\textwidth}
  \centering
  \includegraphics[scale=\figfivescale]{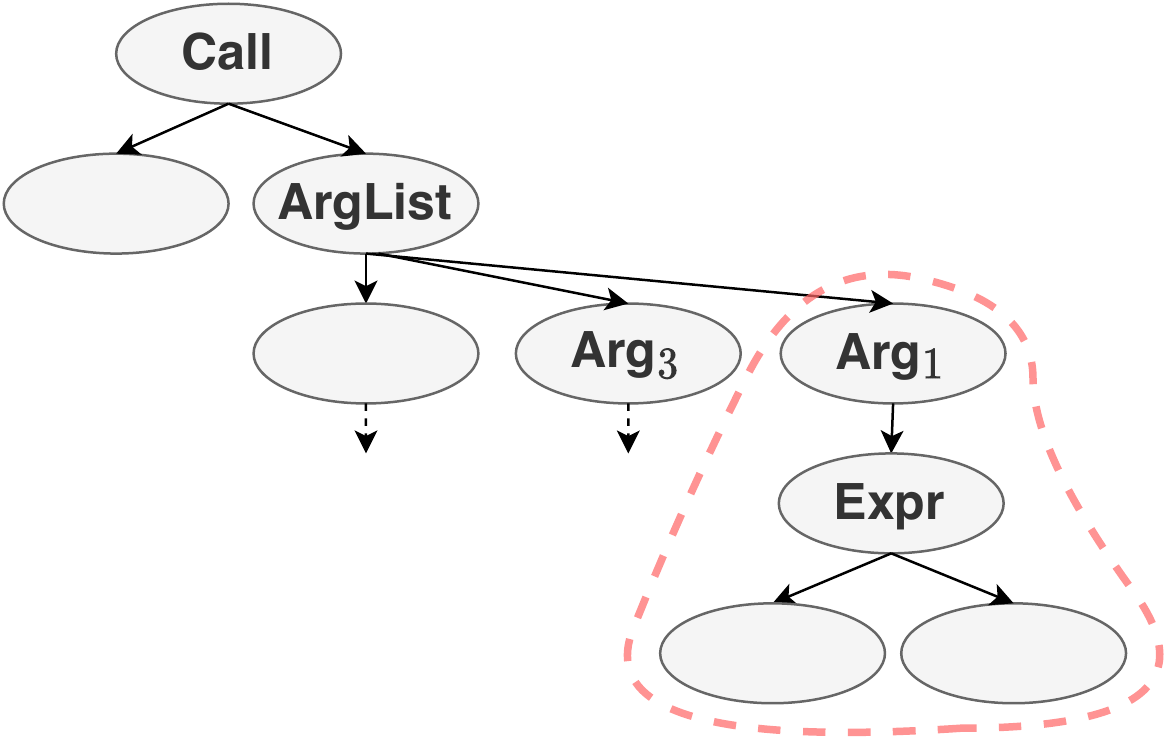}  
  \caption{}
  \label{Fi:mov_exmp_b}
\end{subfigure}

\caption{
An example of a path that represents a \textbf{\scode{MOV}} operation.  \Cref{Fi:mov_exmp_a} shows the path: \scode{Arg$_1$} $\rightarrow$ \scode{ArgList} $\rightarrow$ \scode{Arg$_3$} that represents the edit of moving the first argument to be the last argument. A dotted contour represents the subtree that will be affected by applying the operations. \Cref{Fi:mov_exmp_b} shows the AST after applying the edit.
The affected subtree is surrounded by a dashed contour.}
\label{Fi:editaspath}
\end{figure}

\begin{figure}
    \centering
    \includegraphics[scale=0.5]{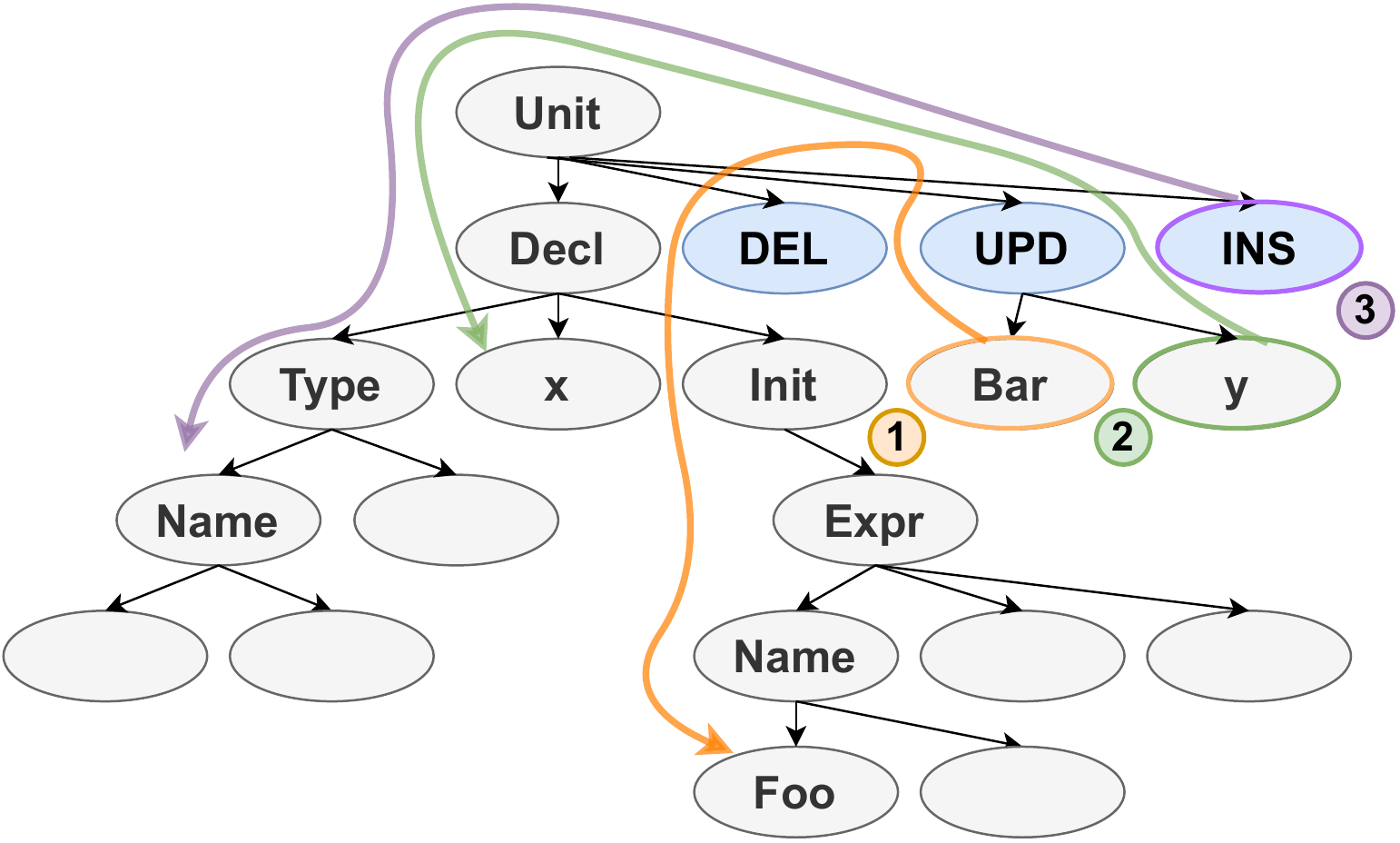}
    \caption{
    An example of \textbf{\scode{UPD}} (update) and \textbf{\scode{INS}} (insert) operations in the \emph{context} $\cbefore$. %
     The \textbf{\textcolor{YellowOrange}{orange\raisebox{.5pt}{ \textcircled{\raisebox{-.9pt} {1}}}}} path represents that the node \scode{Foo} has been updated to the value \scode{Bar}. 
    Similarly, the \textbf{\textcolor{LimeGreen}{green\raisebox{.5pt}{ \textcircled{\raisebox{-.9pt} {2}}}}} path represent that the node \scode{x} has been updated to \scode{y}.
    The \textbf{\textcolor{Plum}{purple\raisebox{.5pt}{ \textcircled{\raisebox{-.9pt} {3}}}}} path represents the insertion of node \scode{Name} along with its subtree.
    }
    \label{Fi:ctx_paths}
\end{figure}

To represent insertions (\textbf{\scode{INS}}) and updates (\textbf{\scode{UPD}}) in the \emph{context} that transformed $\cbefore$ into $\cafter$, we augment the AST with additional \scode{UPD} and \scode{INS} nodes. 
To represent all update operations \textbf{\scode{UPD}} $n_s, n_t$, we add the necessary $n_s$ nodes as children of \scode{UPD}. For example, in \Cref{Fi:ctx_paths}, there are two update operations that involve two source nodes: \scode{y} and \scode{Bar}. Thus, we add these nodes as children of \scode{UPD} and represent the operations with paths that originate from these nodes.
The \textbf{\textcolor{YellowOrange}{orange\raisebox{.5pt}{ \textcircled{\raisebox{-.9pt} {1}}}}} path, for instance, represents the update of \scode{Foo} to become \scode{Bar}. 
In the case of insertion of a new subtree, we represent this operation with a path that originates from \scode{INS} and ends in the root of the subtree. Consider the \textbf{\textcolor{Plum}{purple\raisebox{.5pt}{ \textcircled{\raisebox{-.9pt} {3}}}}} path in \Cref{Fi:ctx_paths}. This path represents that the subtree with the root \scode{Name} was inserted as the leftmost child of \scode{Type}. 
We augment the AST with additional \scode{UPD} and \scode{INS} nodes as additional children of the AST's root, along with the special \scode{DEL} node.

These modifications allow us to represent any edit in the context.
In this work, we focus on edits that can be represented as AST paths in $\pbefore$. 
Examples that require generating code \emph{from scratch} require other, more heavyweight code completion models, and are beyond the scope of this paper.

\section{Model Architecture}\label{Se:Arch}

In this section, we describe our model in detail.
The design of our model is guided by the idea of allowing a neural model to consider multiple edits that occurred in the context ($\Delta_{\mathcal{C}}$), and predict a single path operation that should be applied to $\pbefore$ at every time step. The major challenge is: how do we predict a single path operation? Classifying among a fixed vocabulary of path operations is combinatorially infeasible. Alternatively, decomposing the prediction of a path operation into a sequence of smaller atomic node predictions increases the chances of making a mistake, and can lead to predicting a path operation that is not even valid in the given example. 
We take a different approach. We encode all the path operations that are valid in a given example, and train the model to \emph{point} to a single path operation, only among these valid operations. That is, in every example, the model predicts path operations among a different set of valid operations.

\subsection{High-level View}
\begin{figure}
    \centering
    \includegraphics[scale=1]{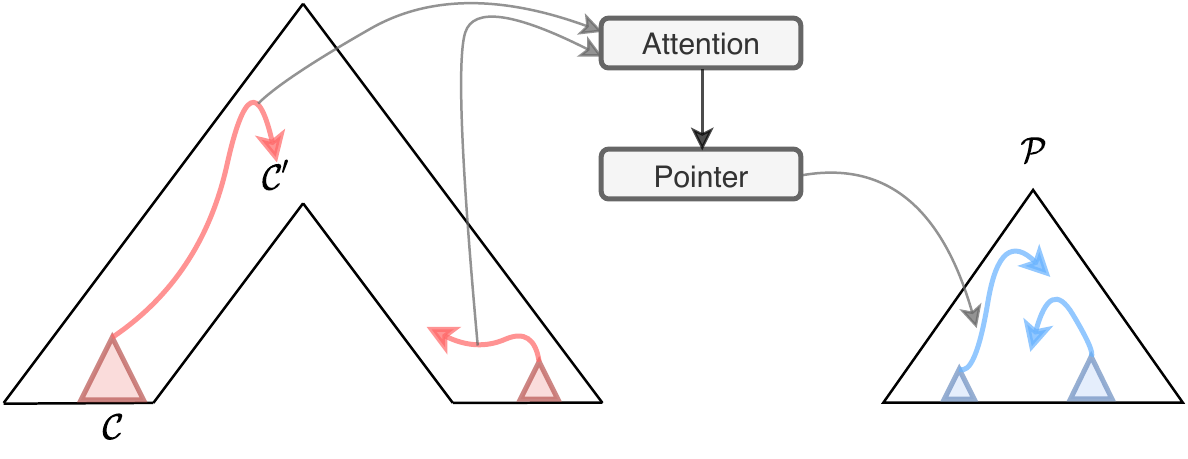}\caption{
    A high-level overview of our architecture. 
    On the left, the partial AST that represents the context $\cbefore$. The red paths represent the transformation from $\cbefore$ to $\cafter$. On the right, we can see the partial AST of $\pbefore$ and its paths that represent possible valid predictions. The model attends to the  paths that transform $\cbefore$ to $\cafter$ to \emph{point} to a path of $\pbefore$ that corresponds to a edit operation.}
    \label{Fi:arch}
\end{figure}

At a high-level, our model reads the edits that occurred in the context and predicts edits that should be performed in the program. Since there may be multiple edits in the context, our model uses \emph{attention} to compute their dynamic weighted average. To predict the edit in the program, our model enumerates all possible edits, expresses them as AST paths, and \emph{points} to the most likely edit.
Thus, the input of the model is a sequence of AST paths from the augmented $\cbefore$, and the output is a sequence of AST paths from $\pbefore$. Our model is illustrated in \Cref{Fi:arch}.

Our model follows the encoder-decoder paradigm. The encoder encodes all valid paths of the input code ($\pbefore$) and the paths of the input context (transforming $\cbefore$ to $\cafter$) into continuous vectors. The decoder generates an edit sequence by pointing to the set of paths in $\pbefore$ while attending to the paths of $\cbefore$ and $\cafter$. 
First, to consider edits that occurred in the context, our model encodes the sequence of context paths that transformed $\cbefore$ into $\cafter$, as a set of vectors. 
Then, the model performs a series of predictions, where each such prediction is an edit that should be applied to $\pbefore$. At each prediction step, the model \emph{attends} (as explained in \Cref{Se:AttnPtr}) to the context paths.
Using the resulting attention vector, the model \emph{points} (as explained in \Cref{Se:Ptr}) to a single path in $\pbefore$. The path in $\pbefore$ that the model points to, is translated to the edit that should be applied in this step. In the next step, the chosen edit from the previous step is used to compute the attention query of the next step.

An edit operation that occurred in the context can be represented as an AST path (\Cref{Se:EditAsPath}). 
We denote the sequence of paths that represent the edits in the context as $\Delta_\mathcal{C}=Paths\left(\cbefore,\cafter\right)$.
The edit function that should be predicted is also represented as AST paths, 
where each path is associated with an edit operation.
We denote the sequence of AST paths that represent the edits that should be predicted as
$\Delta_\mathcal{P}=Paths\left(\pbefore,\pafter\right)$
; we use these vectors as our \emph{classes} from which we make predictions.

Using the above notations, we 
model the conditional probability:
$
    Pr\left(\Delta_\mathcal{P} \mid \Delta_\mathcal{C}\right) %
$.

\subsection{Encoder}
Given a sequence of paths $Paths\left(\cbefore,\cafter\right)$, we encode all paths using a Path Encoder (\Cref{subsub:PathEncoder}).
Then, since it is a sequence that has a meaningful order, the context paths go through an LSTM \cite{hochreiter1997lstm}, resulting in the sequence of vectors $Z_{\mathcal{C}}$.

We enumerate all valid edits that can be applied to $\pbefore$ and denote this set as $Paths\left(\pbefore,\overline{\mathcal{P}}\right)$. We then encode these paths using the Path Encoder (\Cref{subsub:PathEncoder}), which results in the set of vectors $Z_{\overline{\mathcal{P}}}$.
Every path in $Paths\left(\pbefore,\overline{\mathcal{P}}\right)$ can represent different edit operations (\Cref{subsub:OpEncoder}), i.e., both ``update'' and ``move''. Thus, every path vector $z \in Z_{\overline{\mathcal{P}}}$ is projected to represent different edit operations, resulting in the set of vectors $Z_{Op}$, which represent the set of classes from which the model can predict.

\subsubsection{Path Encoder}
\label{subsub:PathEncoder}
Given a set of AST paths, our goal is to create a vector representation $z_i$ for each path $v_{1}...v_{l}$. %
The vocabulary of nodes of the AST is limited to a fixed-size vocabulary from the grammar of the language. In contrast, the \emph{values} of AST leaves correspond to the tokens in the textual representation of the program. Therefore, the vocabulary of these tokens is unbounded.
To address the issue of the unbounded vocabulary of terminal values, we follow previous work \cite{allamanis2015, pmlr-v48-allamanis16, alon2018codeseq}, and split these values into \emph{sub}tokens. For example, the value \scode{toString} will be split into \scode{to} and \scode{string}. 
We represent each path as a sequence of node types using an LSTM, and use subtoken embeddings to represent terminal values (the tokens).

\paragraph{Node Representation}
Each AST path is composed of nodes $v_1,...,v_l$.
Each node is taken from a limited vocabulary of $88$ symbols of the programming language. Terminal nodes also have a user-defined token value.
Every node has an associated child index, i.e., its index among its sibling nodes \cite{Alon_2018}.
We represent each node using a learned embedding matrix $E^{nodes}$ and a learned embedding matrix for its child indices $E^{index}$.
We sum the vector of the node type $w$ with the vector of its child index $i$ to represent the node:
\begin{equation*}
 encode\_node(w) = E^{index}_{i} + E_w^{nodes} \\
\end{equation*}

The first and the last node of an AST path may be terminals whose values are tokens in the code. \footnote{As opposed to code2vec \cite{code2vec}, our paths can originate from and end in nonterminal nodes.}
We use a learned embedding matrix $E^{subtokens}$ to represent each subtoken:
\begin{equation}
    encode\_value(w) = \sum_{s \in split(w)} E_s^{subtokens}
\end{equation}
where $w$ is a value associated with a terminal node.

\paragraph{Path Representation}
We encode the path $v_1,...,v_l$ by applying an LSTM:
\begin{equation*}
h_1,...,h_l = LSTM_{path}(encode\_node\left(v_1\right),...,encode\_node\left(v_l\right))
\end{equation*}
We concatenate the last state vector with an encoding of the values associated with the first and the last nodes in the path, and then pass them through a learned fully connected layer $W_{path}$ and a nonlinearity:
\begin{equation*}
encode\_path\left(v_{1}...v_{l}\right) = tanh \left( W_{path} \cdot \left[ h_l ; encode\_value\left(\phi\left( v_1 \right)\right) ;encode\_value\left(\phi\left( v_l \right)\right) \right] \right)
\end{equation*}

where $\phi$ is the function that retrieves a terminal node's associated value (\Cref{Se:ASTPaths}). If $v_1$ or $v_l$ are nonterminals, and thus do not have an associated value, we encode the first and the last nodes instead of their values; i.e., $encode\_node\left(v\right)$ instead of $encode\_value\left(\phi\left(v\right)\right)$.

To express the order of context paths $Paths\left(\cbefore,\cafter\right)$, we pass these through another LSTM:
\begin{equation}
    Z_{\mathcal{C}} = LSTM_{\mathcal{C}}\left(PathEncoder\left(Paths\left(\cbefore,\cafter\right)\right)\right)
\end{equation}
Applying the path encoder on $Paths\left(\pbefore, \overline{\mathcal{P}}\right)$ results in $Z_{\mathcal{P}}$:
\begin{equation*}
  Z_{\mathcal{P}} = PathEncoder\left(Paths\left(\pbefore, \overline{\mathcal{P}}\right)\right) 
\end{equation*}

\subsubsection{Operation Encoder}
\label{subsub:OpEncoder}
To represent different operations  (i.e., \textbf{\scode{MOV}}, \textbf{\scode{UPD}}, \textbf{\scode{INS}}) that share the same path $z \in Z_{\mathcal{P}}$, we project $z$ using different learned matrices $W_{MOV}, W_{UPD}, W_{INS}$:
\begin{align*}
    z_{MOV} = z\cdot W_{MOV} &&
    z_{UPD} = z\cdot W_{UPD} &&
    z_{INS} = z\cdot W_{INS} 
\end{align*}
such that $z_{MOV}$, $z_{UPD}$, and $z_{INS}$ are used for pointing to \textbf{\scode{MOV}}, \textbf{\scode{UPD}}, and \textbf{\scode{INS}} edits, which are all described by the same encoded path $z$.
This creates our set of possible classes to point to: 
\begin{equation}
    Z_{Op} = \bigcup_{z\in Z_{\mathcal{P}}} \{ z_{MOV}, z_{UPD}, z_{INS} \}
\end{equation}
We use $Z_{Op}$ as the representations of the \emph{classes} over which our model outputs a distribution.

\subsection{Decoder} \label{Se:decoder}
The decoder generates an edit script given the outputs of the encoder. At each decoding time step, the decoder predicts a single edit operation, by pointing to a single vector from $Z_{Op}$, while attending to the sequence of vectors $Z_{\mathcal{C}}$. The decoder consists of three main components: an LSTM \cite{hochreiter1997lstm}, attention \citep{bahdanau2014neural}, and a pointer \citep{vinyals2015pointer}. 

The decoder LSTM operates by receiving an input vector at each time step; then, it uses this input vector to update the LSTM's internal state, and uses the updated state as the query for attention. 
Given the current state, we compute an attention vector of the vectors in $Z_{\mathcal{C}}$, and use the resulting vector to point to a (prediction) vector in $Z_{Op}$.
In the next time step, the input vector for the LSTM is the last pointed to vector from the previous step. %
The initial hidden state of the LSTM is an elementwise average of paths in $Z_{\mathcal{P}}$ and in $Z_{\mathcal{C}}$. %

\paragraph{Attention}
We employ attention as described in \Cref{Se:AttnPtr}, where the query is $h_t$ -- the hidden state of the decoder LSTM at time step $t$.
At each time step, 
we compute a scalar score for every vector $z_i\in Z_{\mathcal{C}}$. This score is computed by performing a dot product between 
each context vector $z_i\in Z_{\mathcal{C}}$ and a learned matrix $W_a$ and $h_t$.  We then normalize all scores with a softmax function to get the normalized weights $\alpha^t$: 
\begin{equation*}
    \alpha^t = \text{softmax}\left( Z_{\mathcal{C}} \cdot W_{a} \cdot h_t^{\top} \right)
\end{equation*}
We then compute a weighted average of $Z_{\mathcal{C}}$ to get the attention vector $\mathbf{c}_t$
\begin{equation*}
    \mathbf{c}_t= \sum_{z_i\in Z_{\mathcal{C}}}^{} \alpha_i \cdot z_i
\end{equation*} 

\paragraph{Pointing}
Given the vector $\mathbf{c}_t$, we compute a pointing score for each valid edit that is represented as $\mathbf{z}\in Z_{Op}$. The resulting scores are normalized using softmax; these normalized scores constitute the model's output distribution. 

We perform a dot product of every $\mathbf{z}\in Z_{Op}$ with another learned weight matrix $W_p$ and $\mathbf{c}_t$. This results in a scalar score  for every valid prediction in $Z_{Op}$.
We then apply a softmax, resulting in a distribution over the vectors in $Z_{Op}$: 
\begin{equation}
    \hat{\mathbf{y}}_t = \text{softmax}\left(Z_{Op} \cdot W_p \cdot \mathbf{c}_t
\right)
\end{equation}
We use this distribution $\hat{\mathbf{y}}_t$ as the model's prediction at time step $t$. At training time, we train all learnable weights to maximize the log-likelihood \cite{rubinstein1999cross}  of $\hat{\mathbf{y}}_t$ according to the true label. 
At test time, we compute the $argmax$ of $\hat{y}_t$ to get the prediction: our model predicts the edit operation that is correlated with the element having the highest pointing score.
The output of the decoder across time steps can be (unambiguously) translated to an edit script.

\section{Experiments}\label{Se:Experiments}

We implemented our approach for \ourtask{} in a neural model called \textbf{C$^{3}$PO}, short for Contextual Code Changes via Path Operations.
The main contributions of our approach are (a) the syntactic representation of code edits; and (b) modeling of the likelihood of code edits, rather than modeling the likelihood of the edited code.
Thus, these are the main ideas that we wish to evaluate.
We compare our model with baselines that represent each of the different paradigms (\Cref{Ti:baselines}) on a new dataset. Our model shows significant performance improvement over the baselines. %

\begin{table}[t]
\small
\begin{tabular}{lccc}
\toprule
\multicolumn{1}{l}{} &\multicolumn{1}{l}{\bf Train} &\multicolumn{1}{l}{\bf Validation} &\multicolumn{1}{l}{\bf Test}
\\ 
\midrule
\# projects                                                &38  &8  &7\\
\# examples                                                &39504  &4468  &5934\\
Avg. number of paths                                       &474  &514  &405\\
Avg. number of edit operations                             &2.6  &2.5  &2.7\\
Avg. number of \textbf{\scode{MOV}}                        &36.4\%  &38.3\%  &41.4\%\\
Avg. number of \textbf{\scode{DEL}}                        &48.6\%  &50\%  &50.8\%\\
Avg. number of \textbf{\scode{INS}}                        &5\%  &4.5\%  &2.8\%\\
Avg. number of \textbf{\scode{UPD}}                        &10.1\%  &7.3\%  &5\%\\
Avg. size of moved subtrees (\textbf{\scode{MOV}})         &3.48  &2.95  &2.85\\
Avg. size of deleted subtrees (\textbf{\scode{DEL}})       &4.49  &4.97  &4.39\\
Avg. size of inserted subtrees (\textbf{\scode{INS}})      &1.27  &2.09  &1.26\\
\bottomrule
\end{tabular}
\caption{Statistics over our dataset.}
\label{Ti:dataset_stats}
\end{table}

\subsection{Dataset}
We introduce a new \ourtask{} dataset of code edits in C\#. We scraped the 53 most popular C\# repositories from GitHub
and extracted all commits since the beginning of the project's history. From each commit, we extracted edits in C\# files along with the edits in their surrounding context. 
Note that a given edit can be part of the edits in the surrounding context ($\Delta_\mathcal{C}$) of one example and can be the edit to be predicted  ($\Delta_\mathcal{P}$) of another example. In other words, the same edit can have different roles in different examples.
We verified that both examples reside in the same split (i.e., either both examples are in the training set, or both examples are in the test set), without leakage between the sets.

For each edit, we considered a context radius of 10 lines,  above and 10 lines below the edit. 
Representing long sequences is computationally difficult for baselines that use Transformers \cite{NIPS2017_7181} because Transformers have a quadratic time and space complexity. We thus limited the context to 10 lines before and after the edit for these baselines. To make a fair comparison, we limited this in our model as well.
We filtered out examples having more than 50 nodes in the AST of $\pbefore$. 
Choosing 50 nodes at most captured the vast majority of examples (81\%). While the technique works for any number of nodes, we picked a limit of 50 to keep the time and cost of experiments reasonable.

To make the task even more challenging, we filtered out examples for which: (a) the edit in $\pbefore$ consists of only \textbf{\scode{DEL}} operations; and (b) edits that both $\pbefore$ and its context contain only \textbf{\scode{UPD}} operations such that all updates in $\mathcal{P}$ are included in the updates of $\mathcal{C}$,
since these usually reflect simple renaming that is easily predicted by modern IDEs.
Following recent work on the adverse effects of code duplication \cite{lopes2017dejavu, allamanis2019adverse}, we split the dataset into training-validation-test \emph{by project}.
This resulted in a dataset containing 39.5K/4.4K/5.9K train/validation/test set examples, respectively. 
We trained all models and baselines on the training set, performed tuning and early-stopping using the validation set, and report final results on the test set.
\Cref{Ti:dataset_stats} shows a summary of the statistics of our dataset.%
 A list of the repositories we used to create our dataset are shown in \Cref{Ap:dataset}. We make our new dataset publicly available at \url{https://github.com/tech-srl/c3po/} .

\subsection{Baselines} \label{baselines}
The two main contributions of our approach that we wish to examine are:
(a) the syntactic representation of code edits; and (b) modeling edit likelihood, rather than modeling code likelihood.
Since we define the new task of \ourtask{}, we picked strong neural baselines and adapted them to this task, to examine the importance of these two main contributions.

\begin{table}[t]
\small
\begin{tabular}{lcc}
\toprule
    & Textual & Syntactic \\
    \midrule
Code Likelihood & \makecell[c]{SequenceR \\ \citep{DBLP:journals/corr/abs-1901-01808}} & \makecell[c]{Path2Tree \\  \cite{aharoni-goldberg-2017-towards}}  \\
\midrule
Edit Likelihood & \makecell[c]{LaserTagger+CRF \\ \cite{malmi2019lasertagger}} & \makecell[c]{\ctc{}PO \\ (this work)} \\
\bottomrule
\end{tabular}
\caption{A high-level taxonomy of our model and the baselines.}
\label{Ti:baselines}
\end{table}

\Cref{Ti:baselines} shows a high-level comparison of our model and the baselines. Each model can be classified across two properties: whether it uses a syntactic or textual representation of the edit, and whether it models the \emph{likelihood of the code} or models the \emph{likelihood of the edit}. 
We put significant effort into performing a fair comparison to all baselines, including subtoken splitting as in our model, lowercasing the subtokens, and replacing generated UNK tokens with the tokens that were given the highest attention score.

\textbf{LaserTagger} \cite{malmi2019lasertagger} - is a textual model that models the \emph{edit likelihood}. LaserTagger learns to apply textual edits to a given text. The model follows the framework of \emph{sequence tagging}, i.e., classifying each token in the input sequence. Each input token is classified into one of: \emph{KEEP$_\phi$}, \emph{DELETE$_\phi$} and \emph{SWAP}, where $\phi$ belongs to a vocabulary of all common phrases obtained from the training set. While LaserTagger leverages edit operations, it does not take advantage of the syntactic structure of the input. 
Since the original implementation of LaserTagger uses a pre-trained BERT NLP model, which cannot be used for code,
we carefully re-implemented a model in their spirit, without BERT. 
We used the same preprocessing scripts and sequence tags as \citet{malmi2019lasertagger}, and encoded the input using either a bidirectional LSTM or a Transformer \cite{NIPS2017_7181} (LaserTagger$_{\text{LSTM}}$ and LaserTagger$_{\text{Transformer}}$, respectively). We further strengthened these models with neural Conditional Random Fields (CRFs) \cite{ma2016end}.
To represent context edits, we employed a sequence alignment algorithm \citep{10.1093/nar/24.14.2730} and extracted the textual edits. We encoded these context edits using 
a bidirectional LSTM
and concatenated the resulting vector to the model's encoded input. %

\textbf{SequenceR} is a re-implementation of \citet{DBLP:journals/corr/abs-1901-01808}.
SequenceR follows the sequence-to-sequence paradigm from Neural Machine Translation (NMT) with attention \citep{DBLP:journals/corr/LuongPM15} and a copy mechanism \cite{DBLP:journals/corr/GuLLL16}. The input is the subtokenized code snippet, along with the textual edits in the context. The output is the edited code. Hence, this method does not take advantage of syntax or edit operations.
We carefully re-implemented this approach because 
SequenceR \emph{abstracts away} identifier names, and replaces identifier names with generic names. For example \scode{int x = 0} becomes \scode{int varInt = 0}. 
Since our model uses identifier names and we found that identifier names help our model, 
to perform a fair comparison -- we kept identifier names in SequenceR as well. 
While the original SequenceR uses LSTMs with copy and attention (SequenceR$_{\text{LSTM}}$), our re-implementation allowed us to strengthen this baseline by replacing
the LSTM with a Transformer \cite{NIPS2017_7181} and a copy mechanism (SequenceR$_{\text{Transformer}}$). We evaluated both SequenceR$_{\text{LSTM}}$, which follows the original model of \citet{DBLP:journals/corr/abs-1901-01808}, and the strengthened SequenceR$_{\text{Transformer}}$ baseline. 

\textbf{Path2Tree} follows \citet{aharoni-goldberg-2017-towards}. This baseline leverages the syntax and models the code likelihood. In this baseline, we performed a pre-order traversal of the AST and represented the AST as a serialized sequence of nodes. Using this sequential serialization of the AST, we could employ strong neural seq2seq models. The input consists of the paths that represent edits in the context (as in our model), along with a serialized sequence that represents the AST of $\pbefore$. The output of the model is the sequence that represents the AST of $\pafter$. 
As the neural underlying seq2seq model, we used both a Transformer (Path2Tree$_{\text{Transformer}}$) with a copy mechanism and a BiLSTM with attention and copy mechanisms (Path2Tree$_{\text{LSTM}}$) .

\subsection{Setup}
From each sample in our dataset, we (a) extracted all paths of 
$Paths\left(\pbefore,\overline{\mathcal{P}}\right)$ that describe possible valid edit operations; and (b) extracted the paths that represent the transformation of $\cbefore$ to $\cafter$, i.e., $Paths\left(\cbefore,\cafter\right)$. We did not filter, discard any of these paths, or limited the paths lengths.

We used input embedding dimensions of 64, LSTM cells with a single layer, and 128 units. %
This resulted in a very lightweight model of only 750K learnable parameters. 
We trained our model on a Tesla V100 GPU using the Adam optimizer \citep{kingma2014method} with a learning rate of 0.001 to minimize the cross-entropy loss. We applied a dropout \citep{DBLP:journals/corr/abs-1207-0580} of $0.25$. %

In the baselines, 
we used BiLSTMs with 2 layers having an embedding and hidden state of size 512; this resulted in 10M learned parameters in SequenceR$_{\text{LSTM}}$ and in Path2Tree$_{\text{LSTM}}$ resulting in 10M learned parameters. We used the original hyperparameters of the Transformer \citep{NIPS2017_7181} to train Transformers in SequenceR$_{\text{Transformer}}$ and Path2Tree$_{\text{Transformer}}$, resulting in 45M learned parameters. 
LaserTagger$_{\text{LSTM}}$ uses BiLSTMs with 2 layers having a hidden state size of 128 and an embedding size of 64. This model contained 1M learned parameters. For LaserTagger$_{\text{Transformer}}$, we used 4 layers of Transformer encoders, with 4 layers and 8 attention heads, an embedding size of 64, and a hidden state size of 512. For both, the context encoders use BiLSTMs with 2 layers having a hidden state size of 128 and an embedding size of 64. We experimented with LaserTaggers 
that contain a context encoder that uses Transformer and setups
that contained larger dimensions, but they achieved slightly lower results. In the other baselines, larger dimensions contributed positively to the performance.

\paragraph{Evaluation Metric}
To perform a fair comparison across all examined models, we had to use a metric that would be meaningful and measurable in all models and baselines. 
We thus measured exact-match accuracy across all models and baselines.
The accuracy of each model is the percentage of examples in the test set for which
the entire target sequence was predicted correctly. 

\subsection{Results}
\begin{table}[t]
\small
\begin{tabular}{lccc}
\toprule
\multicolumn{1}{l}{\bf Model} &\multicolumn{1}{l}{\bf Acc} & \textbf{Learnable Parameters}  & \textbf{Training Time} 
\\ 
\midrule
SequenceR$_{\text{LSTM}}$ \citep{DBLP:journals/corr/abs-1901-01808} +copy &30.7 & 10M & 14h\\
SequenceR$_{\text{Transformer}}$ \citep{DBLP:journals/corr/abs-1901-01808} +copy &32.6  &45M &20h\\
LaserTagger$_{\text{LSTM}}$ \cite{malmi2019lasertagger} +CRF &40.9 & 1M & 10h\\
LaserTagger$_{\text{Transformer}}$ \cite{malmi2019lasertagger} +CRF &41.4 &1.6M & 20h\\
Path2Tree$_{\text{LSTM}}$ \cite{aharoni-goldberg-2017-towards} +copy &22.5 & 10M & 24h\\
Path2Tree$_{\text{Transformer}}$ \cite{aharoni-goldberg-2017-towards} +copy &25.6  & 45M & 24h\\
\midrule
\ctc{}PO (this work)  &\textbf{53.2} & \textbf{750K}  & \textbf{9h}\\
\bottomrule

\end{tabular}
\caption{Our model achieves significantly higher accuracy than the baselines.}
\label{Ti:results}
\end{table}

\paragraph{Performance}
\Cref{Ti:results} depicts the main results of our evaluation: \ctc{}PO gains more than 11\% absolute accuracy over LaserTagger$_{\text{Transformer}}$, which performed the best of all baselines. 
These results emphasize the need for structural representation of both edits and context. \ctc{}PO achieves accuracy that is twice that of the syntactic baseline Path2Tree. Although this baseline uses AST paths to represent the changes in the context of $\pbefore$ and to represent $\pbefore$ with its underlying AST, its performance is inferior compared to our \ctc{}PO. This is because Path2Tree does not model the edit operations directly and thus needs to generate the entire AST of $\pafter$. %

These results show the significance of our model’s two main contributions. \emph{Modeling the edit} has the most significant contribution as expressed in the advantage of our model over both versions of Path2Tree, and in the advantage of both versions of LaserTagger over both versions of SequenceR.
Syntactic representation over textual representation also has a significant contribution, which is expressed in the superiority of our model over both versions of LaserTagger.
Using these two key contributions, our model performs significantly better than all models, while being much more lightweight in terms of learnable parameters. The same results are visualized in \Cref{csharp_results_barchart_figure}.

\definecolor{sequencerlstm}{HTML}{0F9D58}
\definecolor{sequencertrans}{HTML}{000080}
\definecolor{lazerlstm}{HTML}{FF6d00}
\definecolor{lazertrans}{HTML}{46BDC6}
\definecolor{path2treelstm}{HTML}{4285F4}
\definecolor{path2treetrans}{HTML}{DB4437}
\definecolor{ccc}{HTML}{AB30C4}

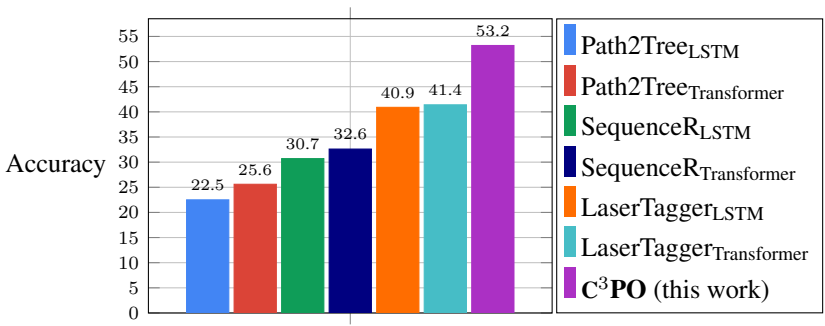
\begin{figure}[h]
\centering
\begin{tikzpicture}
\begin{axis}[
    ybar,
    bar width=16pt,
    width=0.5\textwidth,
    height=5.45cm,
    enlarge x limits=0.10,
    ylabel={Accuracy},
    symbolic x coords={\ },
    xtick=data,
    x tick label style  = {font=\normalsize},
    nodes near coords style={
        color=black,
        font=\tiny
    },%
    nodes near coords,
    font=\scriptsize,
    legend style={font=\normalsize,at={(1.01,0.5)},
    legend cell align={left},
    anchor=west, row sep=0.5pt},
    legend image code/.code={%
            \draw[#1, draw=none] (0cm,-0.1cm) rectangle (0.15cm,0.3cm);
            },
    ytick={0.0,5.0,...,60.0},
    ymin=0,
    grid = major,
    grid style={line width=0.1pt, draw=gray!10},
    major grid style={line width=.2pt,draw=gray!50},
    ylabel style={rotate=-90,font=\normalsize},
    ]
\addplot[path2treelstm,fill=path2treelstm] coordinates {(\ ,22.5)};
\addplot[path2treetrans,fill=path2treetrans] coordinates {(\ ,25.6)};
\addplot[sequencerlstm,fill=sequencerlstm] coordinates {(\ ,30.7)};
\addplot[sequencertrans,fill=sequencertrans] coordinates {(\ ,32.6)};
\addplot[lazerlstm,fill=lazerlstm] coordinates {(\ ,40.9)};
\addplot[lazertrans,fill=lazertrans] coordinates {(\ ,41.4)};
\addplot[ccc,fill=ccc] coordinates {(\ ,53.2)};
\legend{
    Path2Tree$_{\text{LSTM}}$,
    Path2Tree$_{\text{Transformer}}$,
    SequenceR$_{\text{LSTM}}$,
    SequenceR$_{\text{Transformer}}$,
    LaserTagger$_{\text{LSTM}}$,
    LaserTagger$_{\text{Transformer}}$,
    \textbf{\ctc{}PO} (this work)
}
\end{axis}
\end{tikzpicture}
\caption{Visualization of the accuracy score of our model compared to the baselines. The values are the same as in \Cref{Ti:results}. Our model achieves significantly higher
accuracy than the baselines.}
\label{csharp_results_barchart_figure}
\end{figure} 
\definecolor{c3pocolor}{HTML}{AB30C4}

\begin{figure}[t]
\centering
\begin{tikzpicture}[scale=1, trim left=-1.5cm]
	\begin{axis}[
		xlabel={Context radius (lines)},
		ylabel={\footnotesize{Acc}},
		ylabel near ticks,
        xmin=0, xmax=11,
        ymin=-5, ymax=100,
        xtick={1,2,3,4,5,6,7,8,9,10},
        xticklabels={1,2,3,4,5,6,7,8,9,10},
        ytick={0,10,...,100},
        label style={font=\footnotesize},
        ylabel style={rotate=-90,font=\scriptsize},
        ylabel style={font=\scriptsize},
        xlabel style={font=\footnotesize},
        tick label style={font=\scriptsize} ,
        grid = major,
        major grid style={dotted,black},
        width = 0.8\linewidth, height = 5cm,
        name = first
    ]
    \addplot[color=c3pocolor,mark options={fill=c3pocolor, draw=black, line width=0.5pt}, line width=1pt, mark=*, mark size=2pt] coordinates {
    (1,26.7)
    (2,40)
    (3,45.7)
    (4,56)
    (5,56)
    (6,55.9)
    (7,57.6)
    (8,50.8)
    (9,59)
    (10,50.8)
	};
	\end{axis}
	
		\begin{axis}[
		xlabel={Context radius (lines)},
		bar width=16,
		symbolic x coords={0,1,2,3,4,5,6,7,8,9,10,11},
		ylabel={\footnotesize{\# Examples}},
		ylabel near ticks,
        xmin=0, xmax=11,
        xtick=data,
        ytick={0,250,500,...,1750},
        ymin=0,
        ymax=2000,
        label style={font=\footnotesize},
        ylabel style={rotate=-90,font=\scriptsize},
        ylabel style={font=\scriptsize},
        xlabel style={font=\footnotesize},
        tick label style={font=\scriptsize} ,
        grid = major,
        major grid style={dotted,black},
        width = 0.8\linewidth, height = 5cm,
        name=second,
        at=(first.below south west),
        anchor=north west
    ]
        \addplot[ybar,fill=c3pocolor] coordinates { 
        (1,30)
        (2,252)
        (3,212)
        (4,353)
        (5,412)
        (6,602)
        (7,571)
        (8,931)
        (9,884)
        (10,1684)
        };
	\end{axis}

\end{tikzpicture}
\caption{The upper figure depicts the accuracy as a function of the context radius size. The lower figure shows the number of examples per radius size in the test set.}
\label{acc-by-ctx-radius}
\end{figure}
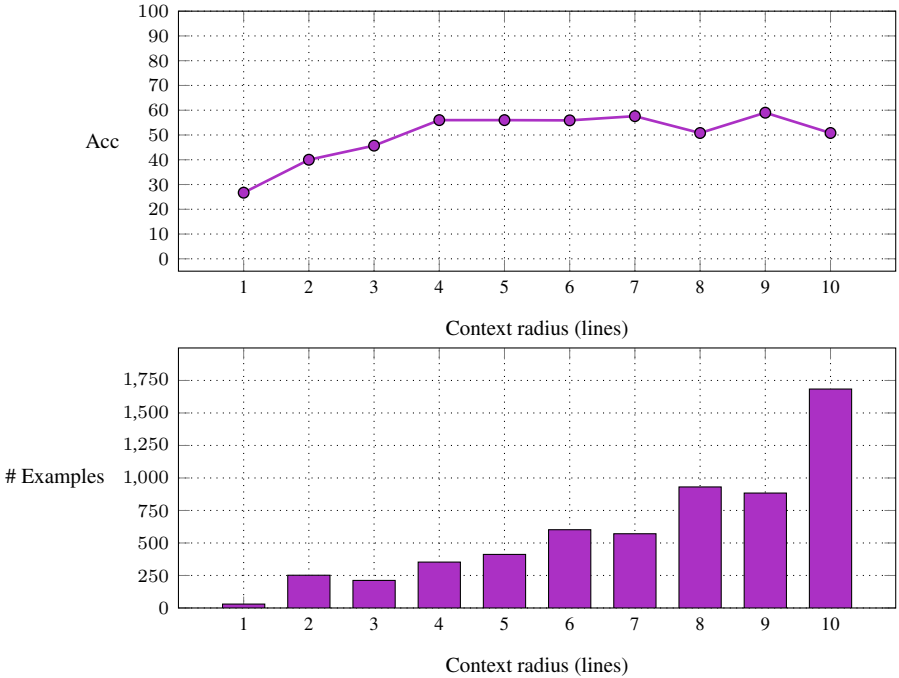 

\paragraph{Path2Tree Lower Performance Compared to SequenceR}
Although Path2Tree represents the edits syntactically (which we believe to be a better representation, in general) and SequenceR represents edits textually, the results of Path2Tree are lower than those of SequenceR.

We believe that the main limitation of Path2Tree is that it cannot easily generalize between the program fragment $\mathcal{P}$ and the given context $\mathcal{C}$. This occurs because $\mathcal{C}$ is represented as paths, while $\mathcal{P'}$ is generated as a tree. In contrast, SequenceR represents all inputs and outputs the same, i.e., as sequences of tokens. We also performed initial experiments with a Tree2Tree baseline that encodes $\mathcal{C}$, $\mathcal{C'}$, and $\mathcal{P}$ as trees and generates $\mathcal{P'}$ as a tree, thus potentially having a better generalization ability than Path2Tree. However, Tree2Tree achieved much lower results than Path2Tree, because the tree encoding created very large inputs, especially in $\mathcal{C}$. These, prevented the model from properly capturing the edits that occurred in the context ($\Delta_{\mathcal{C}}$), while the encoding of $\Delta_{\mathcal{C}}$ as paths is much more succinct and focused (and performed better than Tree2Tree).

\subsection{Scalability Analysis}
We conducted an analysis of our model that shows the performance of \ctc{}PO as a function of the context radius size and the number of nodes.

\Cref{acc-by-ctx-radius} shows the accuracy of \ctc{}PO compared to the context radius size, i.e., the number of lines between the beginning of $\cbefore$ and $\pbefore$.
As shown, the accuracy of \ctc{}PO remains stable when the context radius increases. This hints that the context radius can be further increased without sacrificing accuracy. In our experiments, we put this limitation only to limit the size of the dataset.

\Cref{acc-by-num-nodes} shows the accuracy of \ctc{}PO compared to the number of nodes in the AST of $\pbefore$. As the size of $\pbefore$ increases, our model shows a natural descent, and the accuracy stabilizes for sizes of 31 nodes and above. As shown in the lower part of \Cref{acc-by-num-nodes}, the number of examples also decreases with the size of the edit: the most common edits have 11 to 15 nodes, in which our model achieves an accuracy of 80\%.

\definecolor{c3pocolor}{HTML}{AB30C4}

\begin{figure}[t]
\centering
\begin{tikzpicture}[scale=1, trim left=-1.5cm]
	\begin{axis}[
		xlabel={\# nodes},
		ylabel={\footnotesize{Acc}},
		ylabel near ticks,
        xmin=0, xmax=55,
        ymin=-5, ymax=105,
         xtick={5,10,15,20,25,30,35,40,45,50},
        xticklabels={0--5,6--10,11--15,16--20,21--25,26--30,31--35,36--40,41--45,46--50},
        ytick={0,10,...,100},
        label style={font=\footnotesize},
        ylabel style={rotate=-90,font=\scriptsize},
        ylabel style={font=\scriptsize},
        xlabel style={font=\footnotesize},
        tick label style={font=\scriptsize} ,
        grid = major,
        major grid style={dotted,black},
        width = 0.8\linewidth, height = 5cm,
        name = first
    ]
    \addplot[color=c3pocolor,mark options={fill=c3pocolor, draw=black, line width=0.5pt}, line width=1pt, mark=*, mark size=2pt] coordinates {
    (5,100)
    (10,88.9)
    (15,79.6)
    (20,55.8)
    (25,35.8)
    (30,34.3)
    (35,21)
    (40,22.5)
    (45,21.7)
    (50,22.4)
	};
	\end{axis}
	
		\begin{axis}[
		xlabel={\# nodes},
		bar width=16,
		symbolic x coords={0, 0--5,6--10,11--15,16--20,21--25,26--30,31--35,36--40,41--45,46--50, 55},
		ylabel={\footnotesize{\# Examples}},
		ylabel near ticks,
        ytick={0,250,500,...,1500},
        ymin=0,
        ymax=1750,
        label style={font=\footnotesize},
        ylabel style={rotate=-90,font=\scriptsize},
        ylabel style={font=\scriptsize},
        xlabel style={font=\footnotesize},
        tick label style={font=\scriptsize} ,
        grid = major,
        major grid style={dotted,black},
        width = 0.8\linewidth, height = 5cm,
        name=second,
        at=(first.below south west), 
        anchor=north west
    ]
        \addplot[ybar,fill=c3pocolor] coordinates { 
        (0--5,10)
        (6--10,718)
        (11--15,1498)
        (16--20,919)
        (21--25,772)
        (26--30,655)
        (31--35,496)
        (36--40,462)
        (41--45,267)
        (46--50,138)
        };
	\end{axis}

\end{tikzpicture}
\caption{The upper figure depicts the accuracy as a function of the number of nodes in $\pbefore$. The lower figure shows the number of examples compared to the number of nodes in $\pbefore$.}
\label{acc-by-num-nodes}
\end{figure}
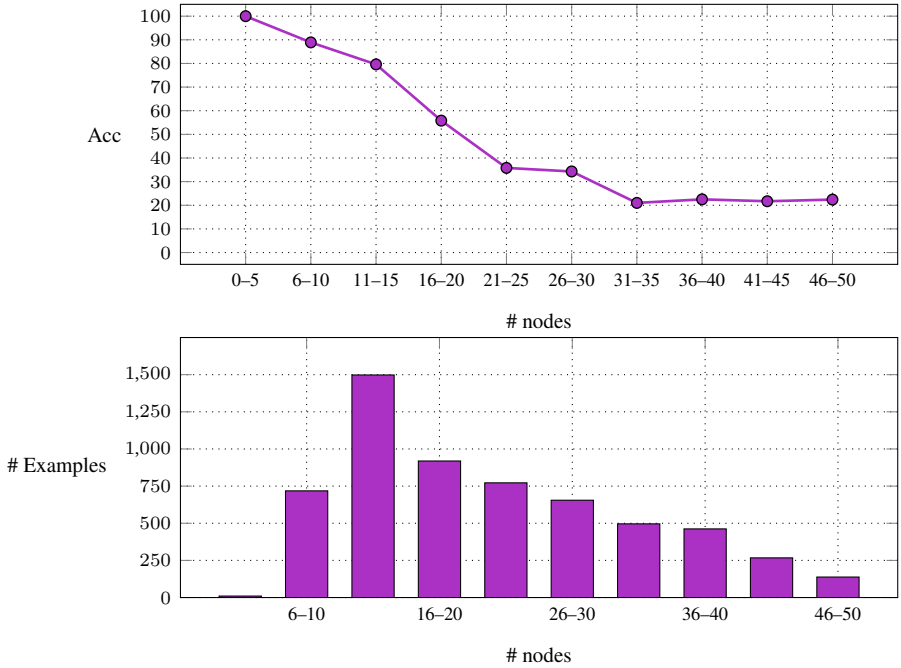 

\subsection{Qualitative Analysis}
We manually examined the predicted examples and discuss two representative cases.

\begin{figure}[t]
\begin{subfigure}[b]{.49\textwidth}
  \centering
  \includegraphics[scale=0.45]{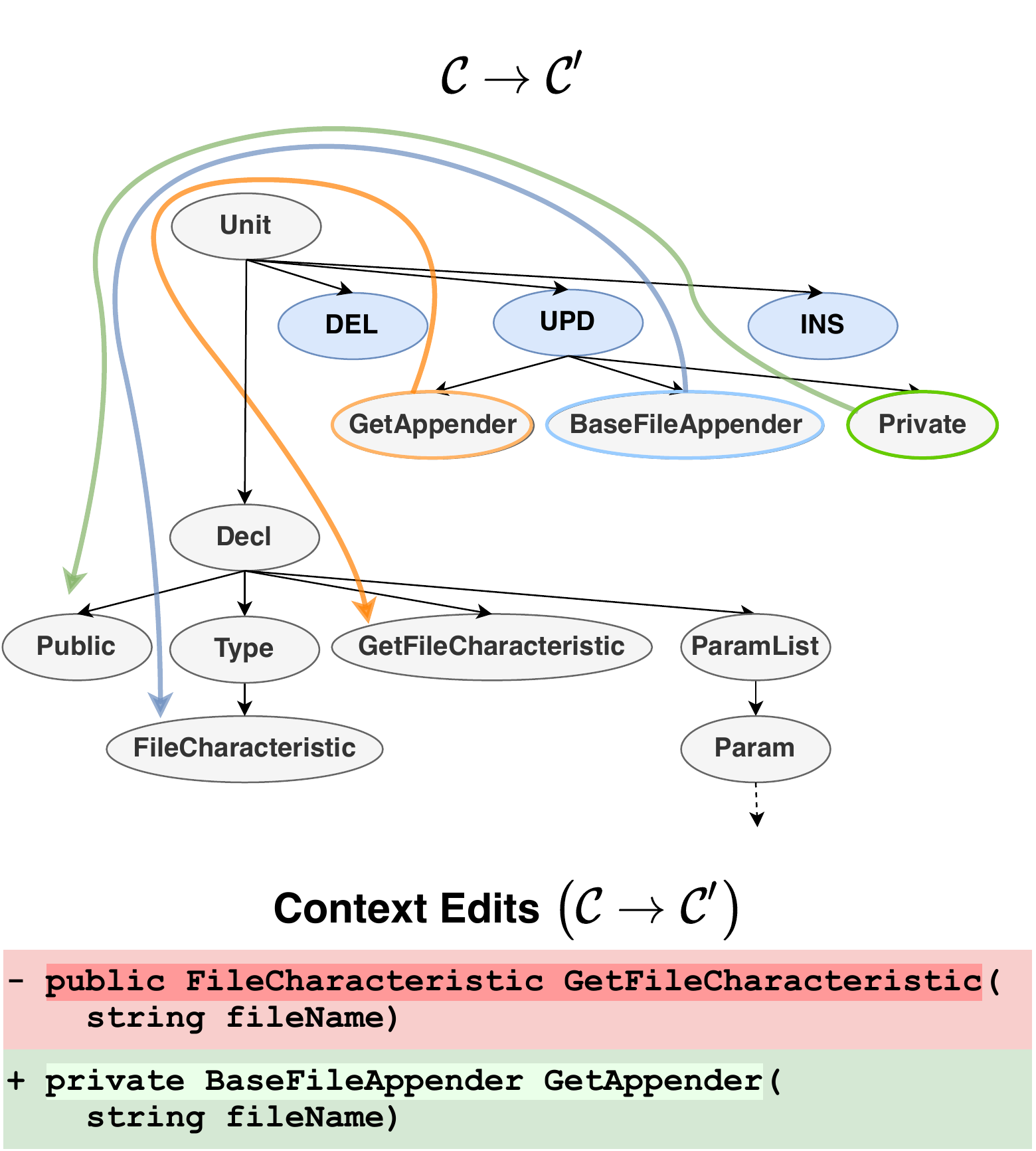}  
  \caption{}
  \label{Fi:qa_mov_exmp_a}
\end{subfigure}
\begin{subfigure}[b]{.49\textwidth}
  \centering
  \includegraphics[scale=0.45]{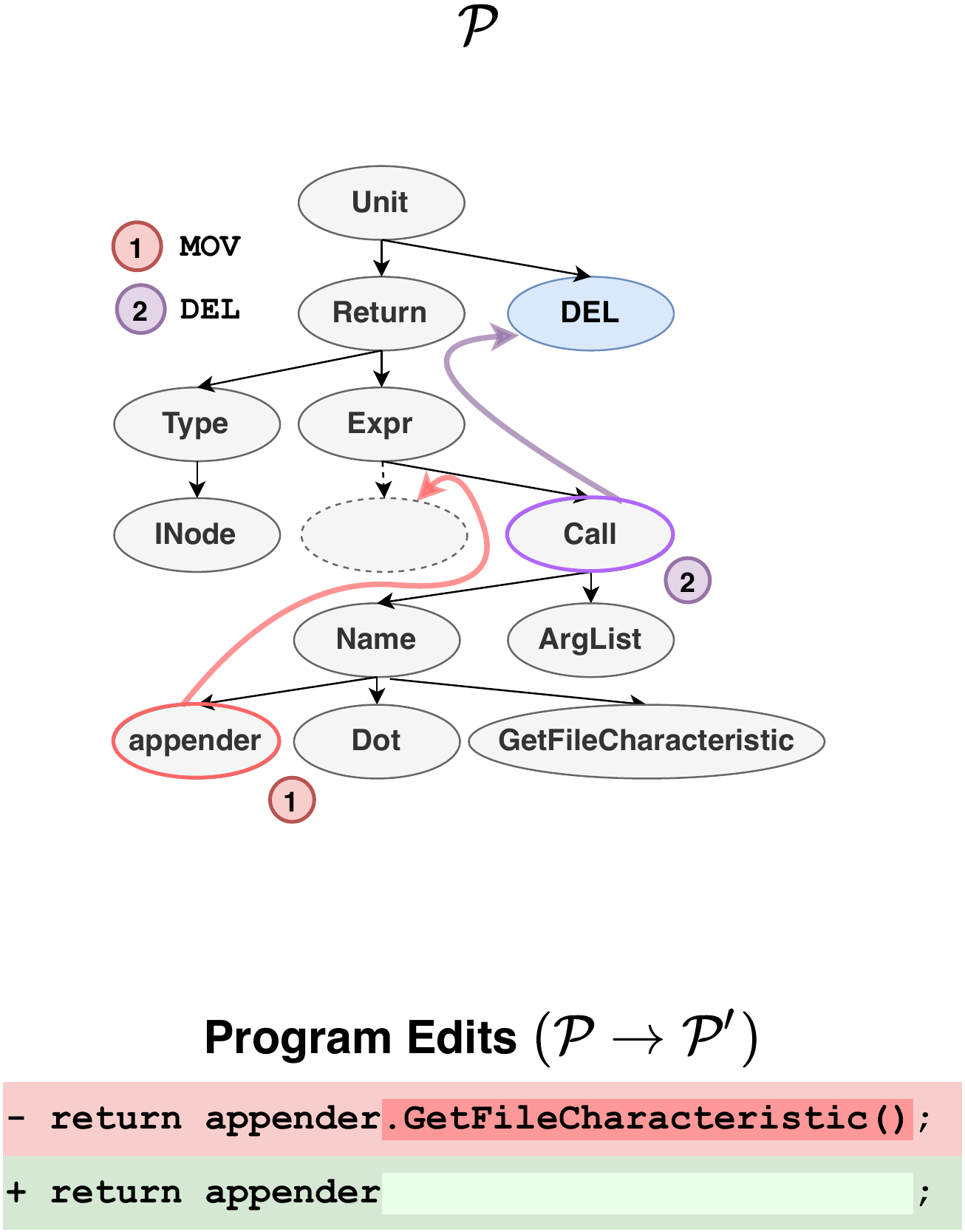}  
  \caption{}
  \label{Fi:qa_mov_exmp_b}
\end{subfigure}
\caption{
An example where the edit of a method signature affects the edit of $\pbefore$ which lies in the method body. \Cref{Fi:qa_mov_exmp_a} illustrates the edit in the context and the paths that describe the transformation from $\cbefore$ to $\cafter$. \Cref{Fi:qa_mov_exmp_b} shows the predicted edit operations along with their associated paths in $\pbefore$.}
\label{Fi:qa_mov_exmp}
\end{figure}

\Cref{Fi:qa_mov_exmp} shows an example in which the modification of a method signature in the context affects $\pbefore$, which lies in the method body.
The context of $\pbefore$, shown in \Cref{Fi:qa_mov_exmp_a}, includes a change in the signature of the method \scode{GetFileCharacteristic}. 
The name of the method was changed to \scode{GetAppender} and its return type was updated from \scode{FileCharacteristic} to \scode{BaseFileAppender}.

Consider $\pbefore$ in \Cref{Fi:qa_mov_exmp_b}. $\pbefore$ is a return statement, located in the body of the changed method \scode{GetFileCharacteristic}. Since the return type of the method was updated to \scode{BaseFileAppender}, the return statements inside the method must be changed as well.  The renaming of the method to \scode{GetAppender} may have also hinted to our model that the \scode{appender} object itself should be returned.
Our model successfully predicted the desirable edit, altering the return statement from \scode{return appender.GetFileCharacteristic} to \scode{return appender;}. 
This example shows how context edits are important in predicting edits in a program, 
by providing information about (a) return type changes and (b) method renaming.

\begin{figure}[t]
\begin{subfigure}[b]{.49\textwidth}
  \centering
  \includegraphics[scale=0.45]{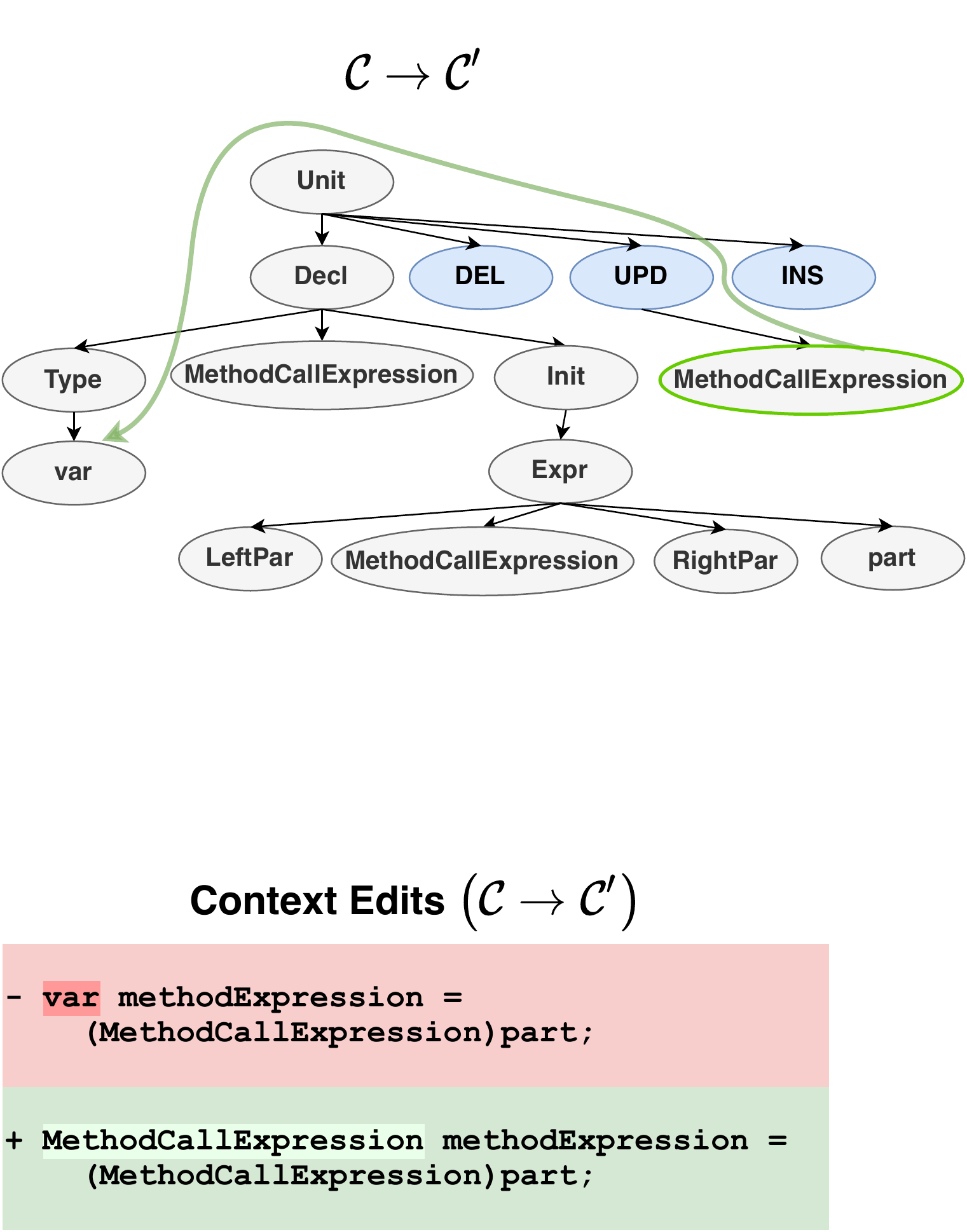}  
  \caption{}
  \label{Fi:qa_ins_exmp_a}
\end{subfigure}
\begin{subfigure}[b]{.49\textwidth}
  \centering
  \includegraphics[scale=0.45]{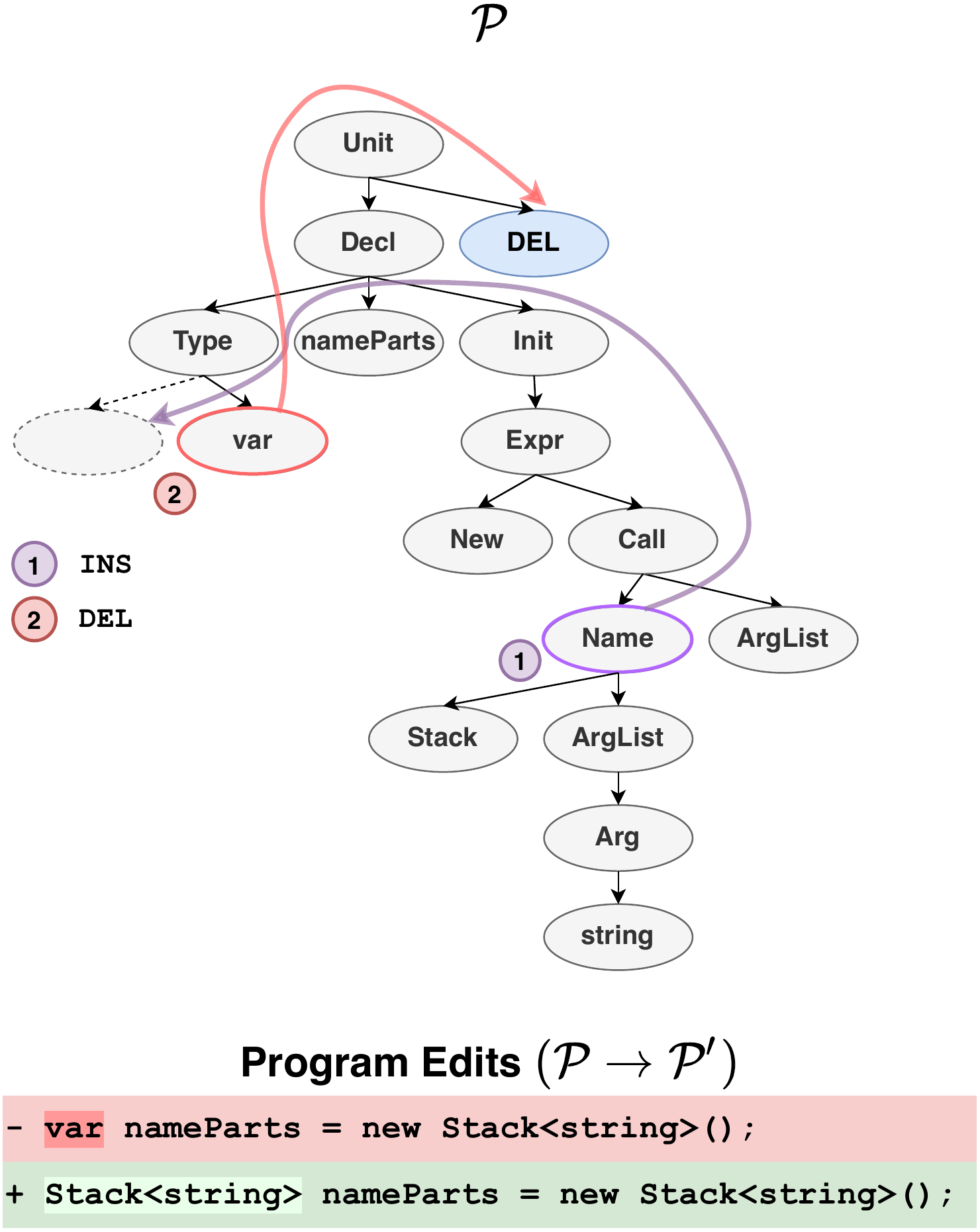}  
  \caption{}
  \label{Fi:qa_ins_exmp_b}
\end{subfigure}
\caption{
An example in which the edit in the context is conceptually similar to the edit of $\pbefore$. \Cref{Fi:qa_ins_exmp_a} illustrates the edit that occurred in the context and the paths that describe the transformation from $\cbefore$ to $\cafter$. \Cref{Fi:qa_ins_exmp_b} shows the predicted edit operations along with their associated paths in $\pbefore$.}
\label{Fi:qa_ins_exmp}
\end{figure}

\Cref{Fi:qa_ins_exmp} illustrates a case where the edit in the context is conceptually similar to the edit in $\pbefore$, but is not identical. 
\Cref{Fi:qa_ins_exmp_a} shows a variable declaration statement, where \scode{part} is cast to the type \scode{MethodCallExpression} and assigned to the newly-declared variable \scode{methodExpression}. In the edited context, the keyword \scode{var} was updated to an explicit type \scode{MethodCallExpression}. 
\Cref{Fi:qa_ins_exmp_b} shows an edit that is similar in spirit: $\pbefore$ consists of an initialization statement, where the variable \scode{nameParts} is  assigned a new \scode{Stack<string>}. Using the edit in the context, our model predicted the edit of \scode{var} to \scode{Stack<string>} in $\pbefore$. This edit consists of an insertion of a new subtree, since \scode{Stack<string>} is represented as a subtree of five nodes. In contrast, the edit in the context is represented as an \textbf{\scode{UPD}} edit, because it only needs to update the value of a single node. 
This example demonstrates a class of examples where the edit in the context hints edits that are similar in spirit in $\pbefore$, but are not identical and should be performed differently.

\section{Ablation Study}\label{Se:Ablation}
We conducted an extensive ablation study to examine the importance of different components in our model. 
We focused on two axes: the representation of $\Delta_{\mathcal{P}}$  and the representation of $\Delta_{\mathcal{C}}$. This allowed us to examine  the origin for the advantage of our model over the strongest baselines, to understand whether it comes from the syntactic representation of the \emph{context} or the syntactic representation of $\pbefore$.

In our model, $\pbefore$ is represented using its syntactic structure, i.e., a path-based representation. Alternatively, $\pbefore$ can be represented using its textual representation. 
The representation of $\pbefore$ determines the representation of $\pafter$. They must be represented similarly, otherwise the model would need to ``translate'' $\pbefore$ into a different representation to predict $\pafter$. However, the representation of the context $\cbefore$ can theoretically be different than that of $\pbefore$.

We thus took our model and examined different representations of the context: path-based context (as in our original model), textual context, and ``no context''.
For each type of context representation, we also experimented with different types of representations for $\pbefore$: syntactic representation, as in our original model, and textual representation of $\pbefore$.
For textual representation of $\pbefore$ we used LaserTagger$_{\text{Transformer}}$ \cite{malmi2019lasertagger}, which we found to be the strongest textual baseline in \Cref{Se:Experiments}. 
All the hybrid models were re-trained, and their performance is shown in \Cref{Ti:ablation}.

\begin{table}[t]
\begin{tabular}{lccc}
\toprule
    & No Context  &  Textual Context & Path-Based Context \\
    \midrule
Textual $\pbefore$ &  \makecell[c]{35.5}  & \makecell[c]{41.4$^\dagger$} & \makecell[c]{39.5} \\
\midrule
\makecell[c]{Path-Based $\pbefore$ \\\ }  & \makecell[c]{46.5 \\\ } & \makecell[c]{48.5 \\\ } & \makecell[c]{\textbf{53.2}$^\dagger$ \\\ctc{} (this work)} \\
\bottomrule
\end{tabular}
\caption{
Variations on our model. 
$\dagger$ marks results that are copied from \Cref{Ti:results}.}
\label{Ti:ablation}
\end{table}

\paragraph{Contribution of Context} 
According to our observations, the contribution of the changes in the context is considerable, for both the textual and path-based representations of $\pbefore$.
Ignoring changes in the context (the left ``No Context'' column of \Cref{Ti:ablation}) results in lower accuracy. This motivates our task of predicting edits \emph{given} the context. Program edits are correlated with edits that occurred in the context and predicting edits should consider the context edits.

\paragraph{$\pbefore$ Representation}
We observed that across all different settings of context representation, a syntactic representation of $\pbefore$ performs better than a textual representation of $\pbefore$. That is, even if the context is textual (the right column of \Cref{Ti:ablation}), a model benefits from a syntactic representation of $\pbefore$.
This advantage is even clearer in the case of ``No context'', where the path-based representation of $\pbefore$ achieves more than 10\% absolute accuracy over the textual representation of $\pbefore$.
A path-based representation of $\pbefore$ allows us to model the edit in $\pbefore$ directly, which makes the learning task much easier and more generalizable.

\paragraph{Context Representation}
As \Cref{Ti:ablation} shows, the representation of the context \emph{should be compatible} with the representation of $\pbefore$. If $\pbefore$ is textual, a textual context performs better; if $\pbefore$ is syntactic, a syntactic context performs better. We hypothesize that matching the context representation to the program representation allows the model to better utilize the context and makes it easier to model the correlation between edits occurring in the context to edits that should be applied to $\pbefore$.

\section{Related Work}\label{Se:relatedWork}

\paragraph{Representing Programs in Learning Models} The representation of programs in learning models is a question that is even more imperative than the learning algorithm we employed or neural architecture. In the last few years, several approaches have been proposed. Early work used the straightforward textual representation, essentially learning from the flat token stream \cite{DBLP:journals/corr/abs-1901-01808, DBLP:journals/corr/abs-1812-08693, vasic2018neural, pmlr-v48-allamanis16, iyer-etal-2016-summarizing}. Although this leverages NLP learning approaches, such models do not leverage the rich syntax of programming languages, and eventually perform worse than other representations, despite their use of strong NLP models. Another line of work represent programs as graphs. These usually augment the AST of a program with additional semantic edges and use a graph neural network to learn from the resulting graph \cite{allamanis2018learning, yin2018learning, brockschmidt2018generative, fernandes2018structured, hellendoorn2020Global}. Graphs provide a natural way to represent programs and allow us to easily augment programs with domain knowledge such as semantic analysis. However, it is unclear how well can these models perform in the absence of full semantic information -- given partial code, given code that cannot be compiled, or languages that are difficult to analyze semantically. 
As in \cite{Alon_2018}, we leverage AST paths to represent programs.
AST paths were shown to be an effective representation for predicting variable names, method names \cite{code2vec}, natural language descriptions \cite{alon2018codeseq}, and code completion \cite{alon2019structural}.
In our task, AST paths allow us to model edits directly, along with the syntactic relationship between the source and the target node of the edits.

\paragraph{Representing Edits}
Much work has been proposed on representing edits. \citet{yin2018learning} proposed a model that learns to apply a given code edit on another given code snippet.
Although this sounds similar to the task we address; with \ourtask{} there is no specific edit in our input that needs to be applied. 
In contrast to \citet{yin2018learning}, the model must predict what should be edited and how, instead of applying a given edit.
In our work, there is no guarantee that the edit that needs to be predicted is included in the context. Furthermore, there could be several edits in the context. Thus, our model needs to choose and predict the most likely edit itself, while the edits that occurred in the context may only be related to the edit that needs to be predicted.

SequenceR \cite{DBLP:journals/corr/abs-1901-01808} used state-of-the-art NMT models to predict bug fixes on single-line buggy programs. Our work is different from their approach when it comes to the representation of the input and the output. \citet{DBLP:journals/corr/abs-1901-01808} represent the code as a token stream, while our approach represents edits as AST paths. Further, their approach attempts to generate the entire edited program, whereas our model models only the edit. We demonstrated the advantage of our approach over SequenceR empirically in \Cref{Se:Experiments}.

One problem connected to ours is the task of fixing compilation errors. \citet{tarlow2019learning} follows the encoder-decoder paradigm, using an encoder that consists of a graph neural network (GNN) that encodes a multi-graph built from the AST and the compilation error log messages. The decoder is a Transformer \cite{NIPS2017_7181} 
that outputs a sequence representing the predicted edit. DeepDelta \cite{48350} used an NMT model in which the input consists of compilation errors and an AST path from the problematic symbol in the code to the root of the tree. The output of their model is a sequence that represents the edit script. 
In our work, pairwise AST paths allow us to model the desired edit directly, instead of predicting an edit using multiple predictions.

Recently, \citet{Dinella2020HOPPITY:} proposed a model called HOPPITY to detect and fix bugs in source code using graph transformations. 
The main difference between our approach and theirs is that  
HOPPITY does not model edit operations directly, as our model does. Rather, it models a graph that represents the input, and uses the resulting node representations to predict actions.
This modeling makes their model predict \emph{unary} edit operations, while our model predicts \emph{binary} edits:
HOPPITY can only predict \emph{single-node edits} in each step, such as deleting a subtree root, inserting a \emph{single node}, and changing a \emph{single node} value. Thus, edits like moving large subtrees require multiple insertion operations of a single node at a time. In our approach, moving and inserting a subtree can be performed by a \emph{single} edit operation. 
\citet{Dinella2020HOPPITY:} evaluated their model on examples that contain \emph{three} single-node operations at most. 
However, as shown in \Cref{Ap:dataset}, the average size of moved subtrees in our train set is 3.48.
Such edits would have required HOPPITY
to generate the entire subtree in the new position (three operations) and delete the subtree in its original place (one operation), resulting in four operations in total. Hence, our average case is larger than the cases examined by HOPPITY.%

CC2Vec \cite{hoang2020cc2vec} represent edits in version-control-systems (e.g., GitHub). However, their approach represents edits \emph{only textually}.
CC2Vec was demonstrated on the tasks of predicting commit messages, predicting bug fixes, and defect prediction; however, their model could not predict the edit itself the way we do in this paper.

\citet{Chakraborty2018Tree2TreeNT} proposed a two-step model that aims to apply edits in code. The first step of their model encodes the sequence that represents a pre-order traversal of the AST of the original code and generates the sequence that represents the AST of the edited code. In the second step, they assign the values of terminal nodes to concrete values. 
Their approach predicts the edit by synthesizing the entire AST. In \Cref{Se:Experiments} we showed the advantage of modeling the likelihood of edits over modeling the likelihood of the code.
Additionally, our model is trained end-to-end, while \citet{Chakraborty2018Tree2TreeNT} trains different components of their model separately.

\section{Conclusion}\label{Se:Conclusion}
We presented a novel approach for representing and predicting edits in code. Our focus is on learning the likelihood of the edit itself, rather than learning the likelihood of the new program.
We use paths from the Abstract Syntax Tree to represent code edits that occurred in a given context, and use these paths to \emph{point} to edits that should be predicted.

We demonstrate the effectiveness of our approach by using the \ourtask{} task to predict edits in a section of code, given edits in its surrounding context. 
We speculate that our direct modeling of the likelihood of edits, and use of the rich structure of code, are the main components that contribute to the strength of our model. We affirm this conjecture in a thorough evaluation and ablation study. Our method performs significantly better than strong neural baselines that leverage syntax but do not model edits directly, or those that model edits but do not leverage syntax.

We believe our approach can serve as a basis for a variety of models and tools that require the modeling and prediction of code edits. Examples include bug fixing, an \ourtask{} assistant in the programmer's IDE, and automatically adapting client code to changes in public external APIs. 
Further, we believe our work can serve as the basis for a future ``neural code reviewer'', to save human effort and time.
To these ends,
we make all our code, dataset, and trained models publicly available at \url{https://github.com/tech-srl/c3po/} .
\newpage
\appendix
\section{Dataset} \label{Ap:dataset}
\Cref{Ti:dataset_repos} lists the GitHub repositories we used to create our dataset. 

\begin{table}[H]
\scriptsize
\begin{tabular}{lll}
\toprule
\multicolumn{1}{l}{\bf Repository} &\multicolumn{1}{l}{\bf User} &\multicolumn{1}{l}{\bf Split}
\\ 
\midrule
corefx	&dotnet & Train\\
shadowsocks-windows	&shadowsocks & Train\\
CodeHub	&CodeHubApp & Train\\
coreclr	&dotnet & Train\\
roslyn	&dotnet & Train\\
PowerShell	&PowerShell & Train\\
WaveFunctionCollapse	&mxgmn & Train\\
SignalR	&SignalR & Train\\
ShareX	&ShareX & Train\\
Nancy	&NancyFx & Train\\
dapper-dot-net	&StackExchange & Train\\
mono	&mono & Train\\
Wox	&Wox-launcher & Train\\
AutoMapper	&AutoMapper & Train\\
RestSharp	&restsharp & Train\\
BotBuilder	&Microsoft & Train\\
SparkleShare	&hbons & Train\\
Newtonsoft.Json	&JamesNK & Train\\
MonoGame	&MonoGame & Train\\
MaterialDesignInXamlToolkit	&MaterialDesignInXAML & Train\\
ReactiveUI	&reactiveui & Train\\
msbuild	&Microsoft & Train\\
aspnetboilerplate	&aspnetboilerplate & Train\\
orleans	&dotnet & Train\\
Hangfire	&HangfireIO & Train\\
Sonarr	&Sonarr & Train\\
dnSpy	&0xd4d & Train\\
Psychson	&brandonlw & Train\\
acat	&intel & Train\\
SpaceEngineers	&KeenSoftwareHouse & Train\\
PushSharp	&Redth & Train\\
cli	&dotnet & Train\\
StackExchange.Redis	&StackExchange & Train\\
akka.net	&akkadotnet & Train\\
framework	&accord-net & Train\\
monodevelop	&mono & Train\\
Opserver	&opserver & Train\\
ravendb	&ravendb & Train\\
OpenLiveWriter	&OpenLiveWriter & Validation\\
Mvc	&aspnet & Validation\\
GVFS	&Microsoft & Validation\\
OpenRA	&OpenRA & Validation\\
Rx.NET	&dotnet & Validation\\
MahApps.Metro	&MahApps & Validation\\
FluentValidation	&JeremySkinner & Validation\\
ILSpy	&icsharpcode & Validation\\
ServiceStack	&ServiceStack & Test\\
choco	&chocolatey & Test\\
duplicati	&duplicati & Test\\
CefSharp	&cefsharp & Test\\
NLog	&NLog & Test\\
JavaScriptServices	&aspnet & Test\\
EntityFrameworkCore	&aspnet  & Test\\
\bottomrule
\end{tabular}
\caption{Our dataset repositories.}
\label{Ti:dataset_repos}
\end{table}

\FloatBarrier
\bibliography{bib}


\begin{thebibliography}{40}


\ifx \showCODEN    \undefined \def \showCODEN     #1{\unskip}     \fi
\ifx \showDOI      \undefined \def \showDOI       #1{#1}\fi
\ifx \showISBNx    \undefined \def \showISBNx     #1{\unskip}     \fi
\ifx \showISBNxiii \undefined \def \showISBNxiii  #1{\unskip}     \fi
\ifx \showISSN     \undefined \def \showISSN      #1{\unskip}     \fi
\ifx \showLCCN     \undefined \def \showLCCN      #1{\unskip}     \fi
\ifx \shownote     \undefined \def \shownote      #1{#1}          \fi
\ifx \showarticletitle \undefined \def \showarticletitle #1{#1}   \fi
\ifx \showURL      \undefined \def \showURL       {\relax}        \fi
\providecommand\bibfield[2]{#2}
\providecommand\bibinfo[2]{#2}
\providecommand\natexlab[1]{#1}
\providecommand\showeprint[2][]{arXiv:#2}

\bibitem[\protect\citeauthoryear{Aharoni and Goldberg}{Aharoni and
  Goldberg}{2017}]%
        {aharoni-goldberg-2017-towards}
\bibfield{author}{\bibinfo{person}{Roee Aharoni} {and} \bibinfo{person}{Yoav
  Goldberg}.} \bibinfo{year}{2017}\natexlab{}.
\newblock \showarticletitle{Towards String-To-Tree Neural Machine Translation}.
  In \bibinfo{booktitle}{\emph{Proceedings of the 55th Annual Meeting of the
  Association for Computational Linguistics (Volume 2: Short Papers)}}.
  \bibinfo{publisher}{Association for Computational Linguistics},
  \bibinfo{address}{Vancouver, Canada}, \bibinfo{pages}{132--140}.
\newblock
\urldef\tempurl%
\url{https://doi.org/10.18653/v1/P17-2021}
\showDOI{\tempurl}


\bibitem[\protect\citeauthoryear{Allamanis}{Allamanis}{2019}]%
        {allamanis2019adverse}
\bibfield{author}{\bibinfo{person}{Miltiadis Allamanis}.}
  \bibinfo{year}{2019}\natexlab{}.
\newblock \showarticletitle{The Adverse Effects of Code Duplication in Machine
  Learning Models of Code}. In \bibinfo{booktitle}{\emph{Proceedings of the
  2019 ACM SIGPLAN International Symposium on New Ideas, New Paradigms, and
  Reflections on Programming and Software}} (Athens, Greece)
  \emph{(\bibinfo{series}{Onward! 2019})}. \bibinfo{publisher}{Association for
  Computing Machinery}, \bibinfo{address}{New York, NY, USA},
  \bibinfo{pages}{143–153}.
\newblock
\showISBNx{9781450369954}
\urldef\tempurl%
\url{https://doi.org/10.1145/3359591.3359735}
\showDOI{\tempurl}


\bibitem[\protect\citeauthoryear{Allamanis, Barr, Bird, and Sutton}{Allamanis
  et~al\mbox{.}}{2015}]%
        {allamanis2015}
\bibfield{author}{\bibinfo{person}{Miltiadis Allamanis},
  \bibinfo{person}{Earl~T. Barr}, \bibinfo{person}{Christian Bird}, {and}
  \bibinfo{person}{Charles Sutton}.} \bibinfo{year}{2015}\natexlab{}.
\newblock \showarticletitle{Suggesting Accurate Method and Class Names}. In
  \bibinfo{booktitle}{\emph{Proceedings of the 2015 10th Joint Meeting on
  Foundations of Software Engineering}} (Bergamo, Italy)
  \emph{(\bibinfo{series}{ESEC/FSE 2015})}. \bibinfo{publisher}{ACM},
  \bibinfo{address}{New York, NY, USA}, \bibinfo{pages}{38--49}.
\newblock
\showISBNx{978-1-4503-3675-8}
\urldef\tempurl%
\url{https://doi.org/10.1145/2786805.2786849}
\showDOI{\tempurl}


\bibitem[\protect\citeauthoryear{Allamanis, Brockschmidt, and
  Khademi}{Allamanis et~al\mbox{.}}{2018}]%
        {allamanis2018learning}
\bibfield{author}{\bibinfo{person}{Miltiadis Allamanis}, \bibinfo{person}{Marc
  Brockschmidt}, {and} \bibinfo{person}{Mahmoud Khademi}.}
  \bibinfo{year}{2018}\natexlab{}.
\newblock \showarticletitle{Learning to Represent Programs with Graphs}. In
  \bibinfo{booktitle}{\emph{International Conference on Learning
  Representations}}.
\newblock
\urldef\tempurl%
\url{https://openreview.net/forum?id=BJOFETxR-}
\showURL{%
\tempurl}


\bibitem[\protect\citeauthoryear{Allamanis, Peng, and Sutton}{Allamanis
  et~al\mbox{.}}{2016}]%
        {pmlr-v48-allamanis16}
\bibfield{author}{\bibinfo{person}{Miltiadis Allamanis}, \bibinfo{person}{Hao
  Peng}, {and} \bibinfo{person}{Charles Sutton}.}
  \bibinfo{year}{2016}\natexlab{}.
\newblock \showarticletitle{A Convolutional Attention Network for Extreme
  Summarization of Source Code}. In \bibinfo{booktitle}{\emph{Proceedings of
  The 33rd International Conference on Machine Learning}}
  \emph{(\bibinfo{series}{Proceedings of Machine Learning Research},
  Vol.~\bibinfo{volume}{48})}, \bibfield{editor}{\bibinfo{person}{Maria~Florina
  Balcan} {and} \bibinfo{person}{Kilian~Q. Weinberger}} (Eds.).
  \bibinfo{publisher}{PMLR}, \bibinfo{address}{New York, New York, USA},
  \bibinfo{pages}{2091--2100}.
\newblock
\urldef\tempurl%
\url{http://proceedings.mlr.press/v48/allamanis16.html}
\showURL{%
\tempurl}


\bibitem[\protect\citeauthoryear{Alon, Brody, Levy, and Yahav}{Alon
  et~al\mbox{.}}{2019a}]%
        {alon2018codeseq}
\bibfield{author}{\bibinfo{person}{Uri Alon}, \bibinfo{person}{Shaked Brody},
  \bibinfo{person}{Omer Levy}, {and} \bibinfo{person}{Eran Yahav}.}
  \bibinfo{year}{2019}\natexlab{a}.
\newblock \showarticletitle{code2seq: Generating Sequences from Structured
  Representations of Code}. In \bibinfo{booktitle}{\emph{International
  Conference on Learning Representations}}.
\newblock
\urldef\tempurl%
\url{https://openreview.net/forum?id=H1gKYo09tX}
\showURL{%
\tempurl}


\bibitem[\protect\citeauthoryear{Alon, Sadaka, Levy, and Yahav}{Alon
  et~al\mbox{.}}{2019b}]%
        {alon2019structural}
\bibfield{author}{\bibinfo{person}{Uri Alon}, \bibinfo{person}{Roy Sadaka},
  \bibinfo{person}{Omer Levy}, {and} \bibinfo{person}{Eran Yahav}.}
  \bibinfo{year}{2019}\natexlab{b}.
\newblock \bibinfo{title}{Structural Language Models of Code}.
\newblock
\newblock
\showeprint[arxiv]{1910.00577}~[cs.LG]


\bibitem[\protect\citeauthoryear{Alon, Zilberstein, Levy, and Yahav}{Alon
  et~al\mbox{.}}{2018}]%
        {Alon_2018}
\bibfield{author}{\bibinfo{person}{Uri Alon}, \bibinfo{person}{Meital
  Zilberstein}, \bibinfo{person}{Omer Levy}, {and} \bibinfo{person}{Eran
  Yahav}.} \bibinfo{year}{2018}\natexlab{}.
\newblock \showarticletitle{A general path-based representation for predicting
  program properties}.
\newblock \bibinfo{journal}{\emph{Proceedings of the 39th ACM SIGPLAN
  Conference on Programming Language Design and Implementation - PLDI 2018}}
  (\bibinfo{year}{2018}).
\newblock
\showISBNx{9781450356985}
\urldef\tempurl%
\url{https://doi.org/10.1145/3192366.3192412}
\showDOI{\tempurl}


\bibitem[\protect\citeauthoryear{Alon, Zilberstein, Levy, and Yahav}{Alon
  et~al\mbox{.}}{2019c}]%
        {code2vec}
\bibfield{author}{\bibinfo{person}{Uri Alon}, \bibinfo{person}{Meital
  Zilberstein}, \bibinfo{person}{Omer Levy}, {and} \bibinfo{person}{Eran
  Yahav}.} \bibinfo{year}{2019}\natexlab{c}.
\newblock \showarticletitle{Code2vec: Learning Distributed Representations of
  Code}.
\newblock \bibinfo{journal}{\emph{Proc. ACM Program. Lang.}}
  \bibinfo{volume}{3}, \bibinfo{number}{POPL}, Article \bibinfo{articleno}{40}
  (\bibinfo{date}{Jan.} \bibinfo{year}{2019}), \bibinfo{numpages}{29}~pages.
\newblock
\urldef\tempurl%
\url{https://doi.org/10.1145/3290353}
\showDOI{\tempurl}


\bibitem[\protect\citeauthoryear{Bahdanau, Cho, and Bengio}{Bahdanau
  et~al\mbox{.}}{2014}]%
        {bahdanau2014neural}
\bibfield{author}{\bibinfo{person}{Dzmitry Bahdanau},
  \bibinfo{person}{Kyunghyun Cho}, {and} \bibinfo{person}{Yoshua Bengio}.}
  \bibinfo{year}{2014}\natexlab{}.
\newblock \showarticletitle{Neural machine translation by jointly learning to
  align and translate}.
\newblock \bibinfo{journal}{\emph{arXiv preprint arXiv:1409.0473}}
  (\bibinfo{year}{2014}).
\newblock


\bibitem[\protect\citeauthoryear{Birney, Thompson, and Gibson}{Birney
  et~al\mbox{.}}{1996}]%
        {10.1093/nar/24.14.2730}
\bibfield{author}{\bibinfo{person}{Ewan Birney}, \bibinfo{person}{Julie~D.
  Thompson}, {and} \bibinfo{person}{Toby~J. Gibson}.}
  \bibinfo{year}{1996}\natexlab{}.
\newblock \showarticletitle{{PairWise and SearchWise: Finding the Optimal
  Alignment in a Simultaneous Comparison of a Protein Profile against All DNA
  Translation Frames}}.
\newblock \bibinfo{journal}{\emph{Nucleic Acids Research}}
  \bibinfo{volume}{24}, \bibinfo{number}{14} (\bibinfo{date}{07}
  \bibinfo{year}{1996}), \bibinfo{pages}{2730--2739}.
\newblock
\showISSN{0305-1048}
\urldef\tempurl%
\url{https://doi.org/10.1093/nar/24.14.2730}
\showDOI{\tempurl}
\showeprint{https://academic.oup.com/nar/article-pdf/24/14/2730/7064078/24-14-2730.pdf}


\bibitem[\protect\citeauthoryear{Brockschmidt, Allamanis, Gaunt, and
  Polozov}{Brockschmidt et~al\mbox{.}}{2019}]%
        {brockschmidt2018generative}
\bibfield{author}{\bibinfo{person}{Marc Brockschmidt},
  \bibinfo{person}{Miltiadis Allamanis}, \bibinfo{person}{Alexander~L. Gaunt},
  {and} \bibinfo{person}{Oleksandr Polozov}.} \bibinfo{year}{2019}\natexlab{}.
\newblock \showarticletitle{Generative Code Modeling with Graphs}. In
  \bibinfo{booktitle}{\emph{International Conference on Learning
  Representations}}.
\newblock
\urldef\tempurl%
\url{https://openreview.net/forum?id=Bke4KsA5FX}
\showURL{%
\tempurl}


\bibitem[\protect\citeauthoryear{Chakraborty, Allamanis, and Ray}{Chakraborty
  et~al\mbox{.}}{2018}]%
        {Chakraborty2018Tree2TreeNT}
\bibfield{author}{\bibinfo{person}{Saikat Chakraborty},
  \bibinfo{person}{Miltiadis Allamanis}, {and} \bibinfo{person}{Baishakhi
  Ray}.} \bibinfo{year}{2018}\natexlab{}.
\newblock \showarticletitle{Tree2Tree Neural Translation Model for Learning
  Source Code Changes}.
\newblock \bibinfo{journal}{\emph{ArXiv}}  \bibinfo{volume}{abs/1810.00314}
  (\bibinfo{year}{2018}).
\newblock


\bibitem[\protect\citeauthoryear{Chan, Jaitly, Le, and Vinyals}{Chan
  et~al\mbox{.}}{2016}]%
        {chan2016listen}
\bibfield{author}{\bibinfo{person}{William Chan}, \bibinfo{person}{Navdeep
  Jaitly}, \bibinfo{person}{Quoc Le}, {and} \bibinfo{person}{Oriol Vinyals}.}
  \bibinfo{year}{2016}\natexlab{}.
\newblock \showarticletitle{Listen, attend and spell: A neural network for
  large vocabulary conversational speech recognition}. In
  \bibinfo{booktitle}{\emph{2016 IEEE International Conference on Acoustics,
  Speech and Signal Processing (ICASSP)}}. IEEE, \bibinfo{pages}{4960--4964}.
\newblock


\bibitem[\protect\citeauthoryear{Chawathe, Rajaraman, Garcia-Molina, and
  Widom}{Chawathe et~al\mbox{.}}{1996}]%
        {10.1145/235968.233366}
\bibfield{author}{\bibinfo{person}{Sudarshan~S. Chawathe},
  \bibinfo{person}{Anand Rajaraman}, \bibinfo{person}{Hector Garcia-Molina},
  {and} \bibinfo{person}{Jennifer Widom}.} \bibinfo{year}{1996}\natexlab{}.
\newblock \showarticletitle{Change Detection in Hierarchically Structured
  Information}.
\newblock \bibinfo{journal}{\emph{SIGMOD Rec.}} \bibinfo{volume}{25},
  \bibinfo{number}{2} (\bibinfo{date}{June} \bibinfo{year}{1996}),
  \bibinfo{pages}{493–504}.
\newblock
\showISSN{0163-5808}
\urldef\tempurl%
\url{https://doi.org/10.1145/235968.233366}
\showDOI{\tempurl}


\bibitem[\protect\citeauthoryear{Chen, Kommrusch, Tufano, Pouchet, Poshyvanyk,
  and Monperrus}{Chen et~al\mbox{.}}{2019}]%
        {DBLP:journals/corr/abs-1901-01808}
\bibfield{author}{\bibinfo{person}{Zimin Chen}, \bibinfo{person}{Steve
  Kommrusch}, \bibinfo{person}{Michele Tufano},
  \bibinfo{person}{Louis{-}No{\"{e}}l Pouchet}, \bibinfo{person}{Denys
  Poshyvanyk}, {and} \bibinfo{person}{Martin Monperrus}.}
  \bibinfo{year}{2019}\natexlab{}.
\newblock \showarticletitle{SequenceR: Sequence-to-Sequence Learning for
  End-to-End Program Repair}.
\newblock \bibinfo{journal}{\emph{CoRR}}  \bibinfo{volume}{abs/1901.01808}
  (\bibinfo{year}{2019}).
\newblock
\showeprint[arxiv]{1901.01808}
\urldef\tempurl%
\url{http://arxiv.org/abs/1901.01808}
\showURL{%
\tempurl}


\bibitem[\protect\citeauthoryear{Dinella, Dai, Li, Naik, Song, and
  Wang}{Dinella et~al\mbox{.}}{2020}]%
        {Dinella2020HOPPITY:}
\bibfield{author}{\bibinfo{person}{Elizabeth Dinella}, \bibinfo{person}{Hanjun
  Dai}, \bibinfo{person}{Ziyang Li}, \bibinfo{person}{Mayur Naik},
  \bibinfo{person}{Le Song}, {and} \bibinfo{person}{Ke Wang}.}
  \bibinfo{year}{2020}\natexlab{}.
\newblock \showarticletitle{HOPPITY: LEARNING GRAPH TRANSFORMATIONS TO DETECT
  AND FIX BUGS IN PROGRAMS}. In \bibinfo{booktitle}{\emph{International
  Conference on Learning Representations}}.
\newblock
\urldef\tempurl%
\url{https://openreview.net/forum?id=SJeqs6EFvB}
\showURL{%
\tempurl}


\bibitem[\protect\citeauthoryear{Falleri, Morandat, Blanc, Martinez, and
  Monperrus}{Falleri et~al\mbox{.}}{2014}]%
        {DBLP:conf/kbse/FalleriMBMM14}
\bibfield{author}{\bibinfo{person}{Jean{-}R{\'{e}}my Falleri},
  \bibinfo{person}{Flor{\'{e}}al Morandat}, \bibinfo{person}{Xavier Blanc},
  \bibinfo{person}{Matias Martinez}, {and} \bibinfo{person}{Martin Monperrus}.}
  \bibinfo{year}{2014}\natexlab{}.
\newblock \showarticletitle{Fine-grained and accurate source code
  differencing}. In \bibinfo{booktitle}{\emph{{ACM/IEEE} International
  Conference on Automated Software Engineering, {ASE} '14, Vasteras, Sweden -
  September 15 - 19, 2014}}. \bibinfo{pages}{313--324}.
\newblock
\urldef\tempurl%
\url{https://doi.org/10.1145/2642937.2642982}
\showDOI{\tempurl}


\bibitem[\protect\citeauthoryear{Fernandes, Allamanis, and
  Brockschmidt}{Fernandes et~al\mbox{.}}{2019}]%
        {fernandes2018structured}
\bibfield{author}{\bibinfo{person}{Patrick Fernandes},
  \bibinfo{person}{Miltiadis Allamanis}, {and} \bibinfo{person}{Marc
  Brockschmidt}.} \bibinfo{year}{2019}\natexlab{}.
\newblock \showarticletitle{Structured Neural Summarization}. In
  \bibinfo{booktitle}{\emph{International Conference on Learning
  Representations}}.
\newblock
\urldef\tempurl%
\url{https://openreview.net/forum?id=H1ersoRqtm}
\showURL{%
\tempurl}


\bibitem[\protect\citeauthoryear{Gu, Lu, Li, and Li}{Gu et~al\mbox{.}}{2016}]%
        {DBLP:journals/corr/GuLLL16}
\bibfield{author}{\bibinfo{person}{Jiatao Gu}, \bibinfo{person}{Zhengdong Lu},
  \bibinfo{person}{Hang Li}, {and} \bibinfo{person}{Victor O.~K. Li}.}
  \bibinfo{year}{2016}\natexlab{}.
\newblock \showarticletitle{Incorporating Copying Mechanism in
  Sequence-to-Sequence Learning}.
\newblock \bibinfo{journal}{\emph{CoRR}}  \bibinfo{volume}{abs/1603.06393}
  (\bibinfo{year}{2016}).
\newblock
\showeprint[arxiv]{1603.06393}
\urldef\tempurl%
\url{http://arxiv.org/abs/1603.06393}
\showURL{%
\tempurl}


\bibitem[\protect\citeauthoryear{Hellendoorn, Sutton, Singh, Maniatis, and
  Bieber}{Hellendoorn et~al\mbox{.}}{2020}]%
        {hellendoorn2020Global}
\bibfield{author}{\bibinfo{person}{Vincent~J. Hellendoorn},
  \bibinfo{person}{Charles Sutton}, \bibinfo{person}{Rishabh Singh},
  \bibinfo{person}{Petros Maniatis}, {and} \bibinfo{person}{David Bieber}.}
  \bibinfo{year}{2020}\natexlab{}.
\newblock \showarticletitle{Global Relational Models of Source Code}. In
  \bibinfo{booktitle}{\emph{International Conference on Learning
  Representations}}.
\newblock
\urldef\tempurl%
\url{https://openreview.net/forum?id=B1lnbRNtwr}
\showURL{%
\tempurl}


\bibitem[\protect\citeauthoryear{Hinton, Srivastava, Krizhevsky, Sutskever, and
  Salakhutdinov}{Hinton et~al\mbox{.}}{2012}]%
        {DBLP:journals/corr/abs-1207-0580}
\bibfield{author}{\bibinfo{person}{Geoffrey~E. Hinton}, \bibinfo{person}{Nitish
  Srivastava}, \bibinfo{person}{Alex Krizhevsky}, \bibinfo{person}{Ilya
  Sutskever}, {and} \bibinfo{person}{Ruslan Salakhutdinov}.}
  \bibinfo{year}{2012}\natexlab{}.
\newblock \showarticletitle{Improving neural networks by preventing
  co-adaptation of feature detectors}.
\newblock \bibinfo{journal}{\emph{CoRR}}  \bibinfo{volume}{abs/1207.0580}
  (\bibinfo{year}{2012}).
\newblock
\showeprint[arxiv]{1207.0580}
\urldef\tempurl%
\url{http://arxiv.org/abs/1207.0580}
\showURL{%
\tempurl}


\bibitem[\protect\citeauthoryear{Hoang, Kang, Lawall, and Lo}{Hoang
  et~al\mbox{.}}{2020}]%
        {hoang2020cc2vec}
\bibfield{author}{\bibinfo{person}{Thong Hoang}, \bibinfo{person}{Hong~Jin
  Kang}, \bibinfo{person}{Julia Lawall}, {and} \bibinfo{person}{David Lo}.}
  \bibinfo{year}{2020}\natexlab{}.
\newblock \bibinfo{title}{CC2Vec: Distributed Representations of Code Changes}.
\newblock
\newblock
\showeprint[arxiv]{2003.05620}~[cs.SE]


\bibitem[\protect\citeauthoryear{Hochreiter and Schmidhuber}{Hochreiter and
  Schmidhuber}{1997}]%
        {hochreiter1997lstm}
\bibfield{author}{\bibinfo{person}{Sepp Hochreiter} {and}
  \bibinfo{person}{J\"{u}rgen Schmidhuber}.} \bibinfo{year}{1997}\natexlab{}.
\newblock \showarticletitle{Long Short-Term Memory}.
\newblock \bibinfo{journal}{\emph{Neural Comput.}} \bibinfo{volume}{9},
  \bibinfo{number}{8} (\bibinfo{date}{Nov.} \bibinfo{year}{1997}),
  \bibinfo{pages}{1735--1780}.
\newblock
\showISSN{0899-7667}
\urldef\tempurl%
\url{https://doi.org/10.1162/neco.1997.9.8.1735}
\showDOI{\tempurl}


\bibitem[\protect\citeauthoryear{Hunt and McIlroy}{Hunt and McIlroy}{1975}]%
        {Hunt1975AnAF}
\bibfield{author}{\bibinfo{person}{James~W. Hunt} {and}
  \bibinfo{person}{M.~Douglas McIlroy}.} \bibinfo{year}{1975}\natexlab{}.
\newblock \showarticletitle{An algorithm for differential file comparison}.
\newblock


\bibitem[\protect\citeauthoryear{Iyer, Konstas, Cheung, and Zettlemoyer}{Iyer
  et~al\mbox{.}}{2016}]%
        {iyer-etal-2016-summarizing}
\bibfield{author}{\bibinfo{person}{Srinivasan Iyer}, \bibinfo{person}{Ioannis
  Konstas}, \bibinfo{person}{Alvin Cheung}, {and} \bibinfo{person}{Luke
  Zettlemoyer}.} \bibinfo{year}{2016}\natexlab{}.
\newblock \showarticletitle{Summarizing Source Code using a Neural Attention
  Model}. In \bibinfo{booktitle}{\emph{Proceedings of the 54th Annual Meeting
  of the Association for Computational Linguistics (Volume 1: Long Papers)}}.
  \bibinfo{publisher}{Association for Computational Linguistics},
  \bibinfo{address}{Berlin, Germany}, \bibinfo{pages}{2073--2083}.
\newblock
\urldef\tempurl%
\url{https://doi.org/10.18653/v1/P16-1195}
\showDOI{\tempurl}


\bibitem[\protect\citeauthoryear{Kingma and Ba}{Kingma and Ba}{2014}]%
        {kingma2014method}
\bibfield{author}{\bibinfo{person}{Diederik~P. Kingma} {and}
  \bibinfo{person}{Jimmy Ba}.} \bibinfo{year}{2014}\natexlab{}.
\newblock \bibinfo{title}{Adam: A Method for Stochastic Optimization}.
\newblock
\newblock
\urldef\tempurl%
\url{http://arxiv.org/abs/1412.6980}
\showURL{%
\tempurl}
\newblock
\shownote{cite arxiv:1412.6980Comment: Published as a conference paper at the
  3rd International Conference for Learning Representations, San Diego, 2015.}


\bibitem[\protect\citeauthoryear{Lopes, Maj, Martins, Saini, Yang, Zitny,
  Sajnani, and Vitek}{Lopes et~al\mbox{.}}{2017}]%
        {lopes2017dejavu}
\bibfield{author}{\bibinfo{person}{Cristina~V Lopes}, \bibinfo{person}{Petr
  Maj}, \bibinfo{person}{Pedro Martins}, \bibinfo{person}{Vaibhav Saini},
  \bibinfo{person}{Di Yang}, \bibinfo{person}{Jakub Zitny},
  \bibinfo{person}{Hitesh Sajnani}, {and} \bibinfo{person}{Jan Vitek}.}
  \bibinfo{year}{2017}\natexlab{}.
\newblock \showarticletitle{D{\'e}j{\`a}Vu: a map of code duplicates on
  GitHub}.
\newblock \bibinfo{journal}{\emph{Proceedings of the ACM on Programming
  Languages}} \bibinfo{volume}{1}, \bibinfo{number}{OOPSLA}
  (\bibinfo{year}{2017}), \bibinfo{pages}{1--28}.
\newblock


\bibitem[\protect\citeauthoryear{Luong, Pham, and Manning}{Luong
  et~al\mbox{.}}{2015}]%
        {DBLP:journals/corr/LuongPM15}
\bibfield{author}{\bibinfo{person}{Minh{-}Thang Luong}, \bibinfo{person}{Hieu
  Pham}, {and} \bibinfo{person}{Christopher~D. Manning}.}
  \bibinfo{year}{2015}\natexlab{}.
\newblock \showarticletitle{Effective Approaches to Attention-based Neural
  Machine Translation}.
\newblock \bibinfo{journal}{\emph{CoRR}}  \bibinfo{volume}{abs/1508.04025}
  (\bibinfo{year}{2015}).
\newblock
\showeprint[arxiv]{1508.04025}
\urldef\tempurl%
\url{http://arxiv.org/abs/1508.04025}
\showURL{%
\tempurl}


\bibitem[\protect\citeauthoryear{Ma and Hovy}{Ma and Hovy}{2016}]%
        {ma2016end}
\bibfield{author}{\bibinfo{person}{Xuezhe Ma} {and} \bibinfo{person}{Eduard
  Hovy}.} \bibinfo{year}{2016}\natexlab{}.
\newblock \showarticletitle{End-to-end Sequence Labeling via Bi-directional
  LSTM-CNNs-CRF}. In \bibinfo{booktitle}{\emph{Proceedings of the 54th Annual
  Meeting of the Association for Computational Linguistics (Volume 1: Long
  Papers)}}. \bibinfo{pages}{1064--1074}.
\newblock


\bibitem[\protect\citeauthoryear{Malmi, Krause, Rothe, Mirylenka, and
  Severyn}{Malmi et~al\mbox{.}}{2019}]%
        {malmi2019lasertagger}
\bibfield{author}{\bibinfo{person}{Eric Malmi}, \bibinfo{person}{Sebastian
  Krause}, \bibinfo{person}{Sascha Rothe}, \bibinfo{person}{Daniil Mirylenka},
  {and} \bibinfo{person}{Aliaksei Severyn}.} \bibinfo{year}{2019}\natexlab{}.
\newblock \showarticletitle{Encode, Tag, Realize: High-Precision Text Editing}.
  In \bibinfo{booktitle}{\emph{EMNLP-IJCNLP}}.
\newblock


\bibitem[\protect\citeauthoryear{Mesbah, Rice, Johnston, Glorioso, and
  Aftandilian}{Mesbah et~al\mbox{.}}{2019}]%
        {48350}
\bibfield{author}{\bibinfo{person}{Ali Mesbah}, \bibinfo{person}{Andrew Rice},
  \bibinfo{person}{Emily Johnston}, \bibinfo{person}{Nick Glorioso}, {and}
  \bibinfo{person}{Edward Aftandilian}.} \bibinfo{year}{2019}\natexlab{}.
\newblock \showarticletitle{DeepDelta: learning to repair compilation errors}.
  In \bibinfo{booktitle}{\emph{Proceedings of the 2019 27th ACM Joint Meeting
  on European Software Engineering Conference and Symposium on the Foundations
  of Software Engineering}}. \bibinfo{pages}{925--936}.
\newblock


\bibitem[\protect\citeauthoryear{Rubinstein}{Rubinstein}{1999}]%
        {rubinstein1999cross}
\bibfield{author}{\bibinfo{person}{Reuven Rubinstein}.}
  \bibinfo{year}{1999}\natexlab{}.
\newblock \showarticletitle{The cross-entropy method for combinatorial and
  continuous optimization}.
\newblock \bibinfo{journal}{\emph{Methodology and Computing in Applied
  Probability}} \bibinfo{volume}{1}, \bibinfo{number}{2}
  (\bibinfo{year}{1999}), \bibinfo{pages}{127--190}.
\newblock


\bibitem[\protect\citeauthoryear{Sutskever, Vinyals, and Le}{Sutskever
  et~al\mbox{.}}{2014}]%
        {DBLP:journals/corr/SutskeverVL14}
\bibfield{author}{\bibinfo{person}{Ilya Sutskever}, \bibinfo{person}{Oriol
  Vinyals}, {and} \bibinfo{person}{Quoc~V. Le}.}
  \bibinfo{year}{2014}\natexlab{}.
\newblock \showarticletitle{Sequence to Sequence Learning with Neural
  Networks}.
\newblock \bibinfo{journal}{\emph{CoRR}}  \bibinfo{volume}{abs/1409.3215}
  (\bibinfo{year}{2014}).
\newblock
\showeprint[arxiv]{1409.3215}
\urldef\tempurl%
\url{http://arxiv.org/abs/1409.3215}
\showURL{%
\tempurl}


\bibitem[\protect\citeauthoryear{Tarlow, Moitra, Rice, Chen, Manzagol, Sutton,
  and Aftandilian}{Tarlow et~al\mbox{.}}{2019}]%
        {tarlow2019learning}
\bibfield{author}{\bibinfo{person}{Daniel Tarlow}, \bibinfo{person}{Subhodeep
  Moitra}, \bibinfo{person}{Andrew Rice}, \bibinfo{person}{Zimin Chen},
  \bibinfo{person}{Pierre-Antoine Manzagol}, \bibinfo{person}{Charles Sutton},
  {and} \bibinfo{person}{Edward Aftandilian}.} \bibinfo{year}{2019}\natexlab{}.
\newblock \bibinfo{title}{Learning to Fix Build Errors with Graph2Diff Neural
  Networks}.
\newblock
\newblock
\showeprint[arxiv]{1911.01205}~[cs.LG]


\bibitem[\protect\citeauthoryear{Tufano, Watson, Bavota, Penta, White, and
  Poshyvanyk}{Tufano et~al\mbox{.}}{2018}]%
        {DBLP:journals/corr/abs-1812-08693}
\bibfield{author}{\bibinfo{person}{Michele Tufano}, \bibinfo{person}{Cody
  Watson}, \bibinfo{person}{Gabriele Bavota}, \bibinfo{person}{Massimiliano~Di
  Penta}, \bibinfo{person}{Martin White}, {and} \bibinfo{person}{Denys
  Poshyvanyk}.} \bibinfo{year}{2018}\natexlab{}.
\newblock \showarticletitle{An Empirical Study on Learning Bug-Fixing Patches
  in the Wild via Neural Machine Translation}.
\newblock \bibinfo{journal}{\emph{CoRR}}  \bibinfo{volume}{abs/1812.08693}
  (\bibinfo{year}{2018}).
\newblock
\showeprint[arxiv]{1812.08693}
\urldef\tempurl%
\url{http://arxiv.org/abs/1812.08693}
\showURL{%
\tempurl}


\bibitem[\protect\citeauthoryear{Vasic, Kanade, Maniatis, Bieber, and
  singh}{Vasic et~al\mbox{.}}{2019}]%
        {vasic2018neural}
\bibfield{author}{\bibinfo{person}{Marko Vasic}, \bibinfo{person}{Aditya
  Kanade}, \bibinfo{person}{Petros Maniatis}, \bibinfo{person}{David Bieber},
  {and} \bibinfo{person}{Rishabh singh}.} \bibinfo{year}{2019}\natexlab{}.
\newblock \showarticletitle{Neural Program Repair by Jointly Learning to
  Localize and Repair}. In \bibinfo{booktitle}{\emph{International Conference
  on Learning Representations}}.
\newblock
\urldef\tempurl%
\url{https://openreview.net/forum?id=ByloJ20qtm}
\showURL{%
\tempurl}


\bibitem[\protect\citeauthoryear{Vaswani, Shazeer, Parmar, Uszkoreit, Jones,
  Gomez, Kaiser, and Polosukhin}{Vaswani et~al\mbox{.}}{2017}]%
        {NIPS2017_7181}
\bibfield{author}{\bibinfo{person}{Ashish Vaswani}, \bibinfo{person}{Noam
  Shazeer}, \bibinfo{person}{Niki Parmar}, \bibinfo{person}{Jakob Uszkoreit},
  \bibinfo{person}{Llion Jones}, \bibinfo{person}{Aidan~N Gomez},
  \bibinfo{person}{\L~ukasz Kaiser}, {and} \bibinfo{person}{Illia Polosukhin}.}
  \bibinfo{year}{2017}\natexlab{}.
\newblock \showarticletitle{Attention is All you Need}.
\newblock In \bibinfo{booktitle}{\emph{Advances in Neural Information
  Processing Systems 30}}, \bibfield{editor}{\bibinfo{person}{I.~Guyon},
  \bibinfo{person}{U.~V. Luxburg}, \bibinfo{person}{S.~Bengio},
  \bibinfo{person}{H.~Wallach}, \bibinfo{person}{R.~Fergus},
  \bibinfo{person}{S.~Vishwanathan}, {and} \bibinfo{person}{R.~Garnett}}
  (Eds.). \bibinfo{publisher}{Curran Associates, Inc.},
  \bibinfo{pages}{5998--6008}.
\newblock
\urldef\tempurl%
\url{http://papers.nips.cc/paper/7181-attention-is-all-you-need.pdf}
\showURL{%
\tempurl}


\bibitem[\protect\citeauthoryear{Vinyals, Fortunato, and Jaitly}{Vinyals
  et~al\mbox{.}}{2015}]%
        {vinyals2015pointer}
\bibfield{author}{\bibinfo{person}{Oriol Vinyals}, \bibinfo{person}{Meire
  Fortunato}, {and} \bibinfo{person}{Navdeep Jaitly}.}
  \bibinfo{year}{2015}\natexlab{}.
\newblock \bibinfo{title}{Pointer Networks}.
\newblock
\newblock
\showeprint[arxiv]{1506.03134}~[stat.ML]


\bibitem[\protect\citeauthoryear{Yin, Neubig, Allamanis, Brockschmidt, and
  Gaunt}{Yin et~al\mbox{.}}{2019}]%
        {yin2018learning}
\bibfield{author}{\bibinfo{person}{Pengcheng Yin}, \bibinfo{person}{Graham
  Neubig}, \bibinfo{person}{Miltiadis Allamanis}, \bibinfo{person}{Marc
  Brockschmidt}, {and} \bibinfo{person}{Alexander~L. Gaunt}.}
  \bibinfo{year}{2019}\natexlab{}.
\newblock \showarticletitle{Learning to Represent Edits}. In
  \bibinfo{booktitle}{\emph{International Conference on Learning
  Representations}}.
\newblock
\urldef\tempurl%
\url{https://openreview.net/forum?id=BJl6AjC5F7}
\showURL{%
\tempurl}


\end{thebibliography}
\clearpage
\end{document}